\documentclass[reprint, amsmath,amssymb, aps, prl ]{revtex4-1}

\usepackage{graphicx}
\usepackage{dcolumn}
\usepackage{bm}
\usepackage{color, diagbox, soul}
\usepackage[usenames, dvipsnames, x11names]{xcolor}
\usepackage[mathscr]{euscript}
\usepackage[colorlinks]{hyperref}
\hypersetup{
 colorlinks=true,
 citecolor=red,
 linkcolor=blue,
 urlcolor=cyan}
 
\newcommand{\ah}{ \hat{a} }
\newcommand{\ad}{ \hat{a}^\dag}
\newcommand{\Ain}{ \hat{A}^\textrm{in} }
\newcommand{\Aind}{ \hat{A}^{\textrm{in}\dag} }
\newcommand{\Aout}{ \hat{A}^\textrm{out} }
\newcommand{\Aoutd}{ \hat{A}^{\textrm{out}\dag} }
\newcommand{\Atel}{ \hat{A}^\textrm{tel} }
\newcommand{\Ateld}{ \hat{A}^{\textrm{tel}\dag} }

\newcommand{\tah}{ \hat{\alpha} }
\newcommand{\tad}{ \hat{\alpha}^\dag}
\newcommand{\tAin}{ \hat{\mathcal{A}}^\textrm{in} }
\newcommand{\tAind}{ \hat{\mathcal{A}}^{\textrm{in}\dag} }
\newcommand{\bA}{ \hat{\mathbb{A}}}
\newcommand{\bbA}{ \hat{\bar{\mathbb{A}}}}

\newcommand{\hJ}{ \hat{J}}
\newcommand{\sG}{ \mathcal{G}}
\newcommand{\sT}{ \mathcal{T}}
\newcommand{\sF}{ \bm{\mathcal{F}}}
\newcommand{\sR}{ \mathcal{R}}
\newcommand{\hI}{ \bm{I}}

\newcommand{\Chi}{ \mathcal{X}}

\newcommand{\w}{ \omega}

\newcommand{\tD}{ \tilde{\Delta}}
\newcommand{\Wm}{\Omega_\textrm{m}}
\newcommand{\tm}{\textrm{m}}

\begin{document}

\title{Ground state cooling and high-fidelity quantum transduction via parametrically-driven bad-cavity optomechanics}

\author{Hoi-Kwan Lau}
 \email{hklau.physics@gmail.com}
\author{Aashish A. Clerk}

\affiliation{Institute for Molecular Engineering, 
University of Chicago, 
5640 South Ellis Avenue, 
Chicago, Illinois 60637, U.S.A.}

\date{\today}

\begin{abstract}
Optomechanical couplings involve both beam-splitter and two-mode-squeezing types of interactions.  While the former underlies the utility of many applications, the latter creates unwanted excitations and is usually detrimental.  In this work, we propose a simple but powerful method based on cavity parametric driving to suppress the unwanted excitation that does not require working with a deeply sideband-resolved cavity.  Our approach is based on a simple observation:  as both the optomechanical two-mode-squeezing interaction and the cavity parametric drive induce squeezing transformations of the relevant photonic bath modes, they can be made to cancel one another.  We illustrate how our method can cool a mechanical oscillator below the quantum back-action limit, and significantly suppress the output noise of a sideband-unresolved optomechanical transducer.  
\end{abstract}

\maketitle

\textit{Introduction--}  
Optomechanical systems couple mechanical to electromagnetic degrees of freedom, and have a wide range of utility in 
both classical and quantum settings \cite{1995PhRvA..51.2537L, 2014RvMP...86.1391A}.  
Experiments almost always employ a strong electromagnetic drive, with the resulting optomechanical coupling containing both a beam-splitter (BS) interaction and a two-mode-squeezing interaction (TMSI).  The BS interaction exchanges phononic and photonic excitations, and underlies the functionality of numerous optomechanical applications.  This includes cavity cooling of a mechanical mode, 
where mechanical excitations are transferred to the dissipative electromagnetic cavity 
\cite{Teufel:2011jg, Chan:2011dy}.  It also includes the application of quantum transduction:  the mechanical oscillator and BS interaction can be used to mediate microwave to optics quantum state transfer \cite{2011NJPh...13a3017S, 2011JPhCS.264a2025R, 2014NatPh..10..321A, Higginbotham:2018ca}, which is crucial for distributed quantum information processing \cite{2011JPhCS.264a2025R, 2015AnP...527....1T, 2015PNAS..112.3866K}.

In these applications, the TMSI, which simultaneously creates both motional and photonic excitations, is highly undesirable.  The standard strategy is to partially suppress TMSI by making it highly non-resonant, via an appropriate choice of drive frequency and the use of 
low-loss cavities whose damping rate  $\kappa$ is much smaller than the mechanical frequency $\Omega_{\rm m}$.    
This however places stringent restrictions on experimental platforms.  For cavity cooling, 
the residual TMSI sets the fundamental quantum backaction limit on the lowest achievable mechanical occupancy \cite{Marquardt:2007dn,WilsonRae:2007jp}.  This limit prevents approaching the quantum ground state for sideband unresolved systems having $\kappa \gtrsim \Omega_{\rm m}$.  
Strategies for ground state cooling have been formulated for the latter
systems (e.g. dissipative coupling \cite{2009PhRvL.102t7209E, 2011PhRvL.107u3604X, 2015PhRvL.114d3601S, Huang:2018br}, coupling to trapped atoms \cite{2009PhRvA..80f1803G, 2011PhRvA..84e1801G, Lau:2018ev}, photonic squeezing \cite{2009PhRvA..79a3821H, 2018PhyE..102...83A, 2017Natur.541..191C, 2016PhRvA..94e1801A}); however, they present their
 own implementation challenges.
Similarly, TMSI makes high fidelity quantum transduction impossible in the sideband unresolved systems, and even constrains the performance of sideband resolved systems \cite{2014NatPh..10..321A, Higginbotham:2018ca}.
Therefore, sideband-resolving cavity is widely believed to be necessary for efficient transduction \cite{2015AnP...527....1T}, and we are not aware of any correction strategy.

\begin{figure}[t]
\begin{center}
\includegraphics[width=\linewidth]{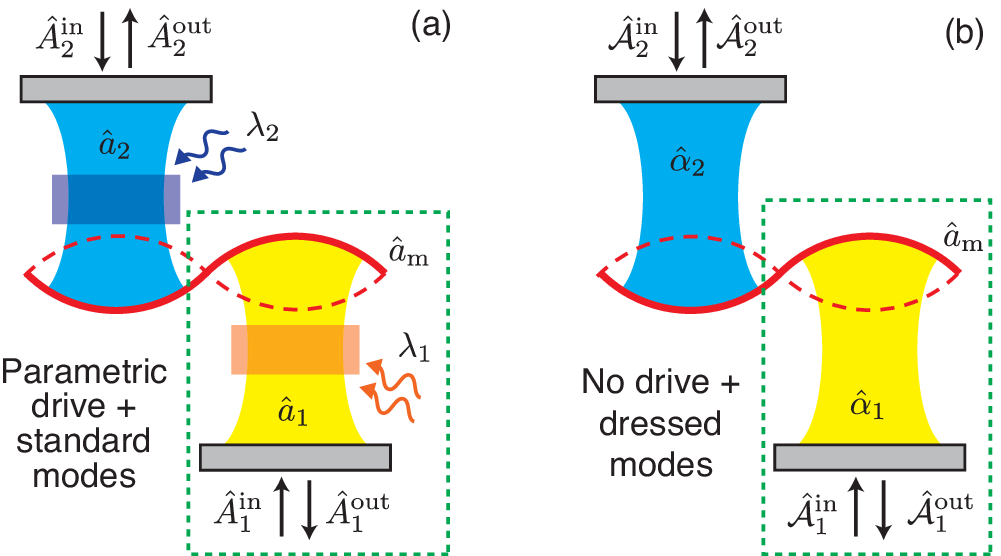}
\caption{(a) Generic optomechanical system described by Eq.~(\ref{eq:LE_lab}), where two parametrically driven cavities ($\hat{a}_1$, $\hat{a}_2$) couple to a common mechanical oscillator ($\hat{a}_\textrm{m}$).  (b) An equivalent optomechanical system described by Eq.~(\ref{eq:LE_eff}), which is not parametrically driven but where both the bath and cavity modes are squeezed.  
Both cavity modes $1$ and $2$ are involved in transduction, but only cavity mode $1$ is utilized in cooling (inside green dotted box).
\label{fig:Intro}}
\end{center}
\end{figure}

In this work, we propose simple but powerful strategies to suppress the deleterious effects in sideband-unresolved optomechanics applications.
Our strategy is based on two key observations. First, 
unwanted backaction effect arises because the mechanical oscillator is driven by electromagnetic vacuum noise that is squeezed by the TMSI.   Second, parametric (i.e. two-photon) driving induces additional electromagnetic squeezing, so that the net squeezing of the cavity backaction can be minimized by optimizing system parameters.  Remarkably, our strategy can completely eliminate TMSI imperfections in sideband unresolved systems in the typical operation regime of large co-operativity but no strong coupling.  
In the case of cooling, our strategy completely eliminates backaction heating.  
In the case of transduction, parametric drive removes the unwanted amplification of the input signal.  With this improvement, we are also able to obtain a new impedance-matching condition and injected squeezing strategy, which both improve the transmission bandwidth and suppress the added noise of the output signal.

\textit{Parametrically driven optomechanics--}  We consider a generic optomechanical system where two cavities ($\ah_1$ and $\ah_2$) are coupled to the same mechanical oscillator ($\ah_\textrm{m}$) (see Fig.~\ref{fig:Intro}a); each cavity is subject to both a linear drive and a parametric drive (PD),  
and the drive frequencies for each cavity are commensurate.  After making standard displacement and linearization transformations  \cite{2014RvMP...86.1391A}, the mode dynamics obey the Langevin equations
\begin{eqnarray}\label{eq:LE_lab}
\dot{\ah}_i &=& -i (\Delta_i-i\frac{\kappa_i}{2}) \ah_i  -i \lambda_i \ad_i -i G_i (\ah_\textrm{m} + \ad_\textrm{m}) -i \sqrt{\kappa_i} \Ain_i \nonumber \\
\dot{\ah}_\textrm{m} &=& -i (\Wm-i\frac{\gamma}{2}) \ah_\textrm{m} -i \sum_{i=1,2}G_i (\ah_i + \ad_i) -i \sqrt{\gamma} \Ain_\textrm{m} ~,
\end{eqnarray}
or in a compact form $\dot{\vec{a}} = -i \bm{H} \vec{a} + \bm{K} \vec{A}^\textrm{in}$, where 
we index vectors and matrices according to the mode operators, i.e.  for any vector $\vec{v}\equiv (v_1, v_{1^\dag}, v_\textrm{m}, v_{\textrm{m}^\dag}, v_2, v_{2^\dag})^\textrm{T}$, and similarly for matrices;  $\bm{K}\equiv \textrm{diag}(-i\sqrt{\kappa_1}, i\sqrt{\kappa_1}, -i\sqrt{\gamma}, i\sqrt{\gamma}, -i\sqrt{\kappa_2}, i\sqrt{\kappa_2})$.  $\Ain_i$ and $\Ain_\textrm{m}$ respectively describe the incident noise on cavity $i$ and the mechanical oscillator; we take $\Ain_i$ to correspond to vacuum noise unless specified otherwise.  The dynamical matrix $\bm{H}$ contains all system parameters: $\Delta_i$, $\lambda_i$, $\kappa_i$, and $G_i$ are the mode-drive frequency detuning, PD strength, dissipation rate, and many-photon optomechanical coupling strength of cavity $i$; $\Wm$ and $\gamma$ are the frequency and damping rate of mechanical oscillator.

We focus attention on parameters where the detunings are large enough that the system is dynamically stable when $G_i=0$ and even without dissipation, i.e. $|\lambda_i| <|\Delta_i|$.  
More discussion about the general system stability is given in Supplementary Material.
The Hamiltonian of each isolated (but parametrically driven) cavity can be diagonalized in terms of a dressed (Bogoliubov) mode:  
$\tah_i = e^{i \phi_i}\cosh r_i \ah_i + e^{i \phi_i} e^{i \theta_i}\sinh r_i \ad_i$, for $i \in \{1,2\}$.  The transformation parameters are: 
\begin{equation}\label{eq:Bogo_para}
e^{i\theta_i}\tanh 2 r_i \equiv \lambda_i/\Delta_i ~~,~~ e^{i\phi_i} \equiv \mu_i/|\mu_i| ~,
\end{equation}
where $\mu_i \equiv \cosh r_i - e^{i\theta_i}\sinh r_i$.  The evolution of the dressed modes follow
\begin{eqnarray}\label{eq:LE_eff}
\dot{\tah}_i &=& -i (\tD_i-i\frac{\kappa_i}{2}) \tah_i -i \sG_i (\ah_\textrm{m} + \ad_\textrm{m}) -i \sqrt{\kappa_i} \tAin_i  \\
\dot{\ah}_\textrm{m} &=& -i (\Wm - i\frac{\gamma}{2}) \ah_\textrm{m}   -i \sum_{i=1,2}\sG_i (\tah_i + \tad_i) -i \sqrt{\gamma} \Ain_\textrm{m}~, \nonumber
\end{eqnarray}
or in the compact form $\dot{\vec{\alpha}}=-i\bm{\mathcal{H}}\vec{\alpha}+\bm{K}\vec{\mathcal{A}}^\textrm{in}$ \footnote{Note that the mechanical mode and bath are unaffected by the diagonalization, i.e. $\tah_\textrm{m} = \ah_\textrm{m}$ and $\tAin_\textrm{m} = \Ain_\textrm{m}$.} .  The dressed modes have modified detunings and optomechanical coupling as $\tD_i \equiv \sqrt{\Delta_i^2 - |\lambda_i|^2}>0$ and $\sG_i \equiv |\mu_i| G_i $, but their dissipation rates remain $\kappa_i$.  
Hereafter, we implicitly assume weak coupling and low mechanical dampling, i.e. $\gamma \ll \sG_1, \sG_2 \ll \tD_i, \kappa_i$, which is the typical experimental regime.

The noise-free dynamics of the dressed modes in Eq.~(\ref{eq:LE_eff}) 
corresponds to a standard optomechanical system with no PD.
This structure results from the optomechanical coupling being a product of quadratures, a structure
that is preserved after a Bogoliubov transformation \cite{Weedbrook:2012fe}. The main difference in Eq.~(\ref{eq:LE_eff}) is in the noise terms:
the input noise from the cavity baths now appears squeezed:
\begin{equation}\label{eq:Bogo_bath}
\tAin_i \equiv  e^{i\phi_i}\cosh r_i \Ain_i -e^{i\phi_i} e^{i\theta_i}\sinh r_i \Aind_i~.
\end{equation}
Hence, the PDs have allowed us to map our system (which is driven by vacuum noise) to a standard optomechanical system driven by squeezed noise.

As we will see, TMSI is detrimental because it induces unwanted squeezing of the bath fluctuations; this squeezing is large when $\kappa \gtrsim \Omega_{\rm m}$.
To correct such effects, our systematic strategy is to use PD to counteract this squeezing.  We first determine the PD parameters needed for this compensation (c.f. Eq.~(\ref{eq:Bogo_bath})) while keeping the dressed-mode dynamics (as given by $\bm{\mathcal{H}}$) fixed.  Next, we optimize $\bm{\mathcal{H}}$ for a specific application.  Finally, the experimentally relevant ``bare" system parameters (given by $\bm{H}$) can be inferred.

\textit{Optomechanical cooling--}  We first apply our strategy to suppress the TMSI-induced backaction heating in optomechanical cooling.  We consider the standard cooling setup involving only one cavity, and thus decouple cavity 2 by setting $G_2=0$.  
The mechanical steady state is determined by its response to the various input noise operators \cite{Marquardt:2007dn}:
$\ah_\textrm{m}[\w] = \bA_1(\w) + \bA_\textrm{m}(\w)$,
where $\hat{O}[\w]\equiv \int \hat{O}(t) e^{i\w t}dt$.  $\bA_1(\w)$ ($\bA_\textrm{m}(\w)$) contains only the photonic (mechanical) bath operators, and thus corresponds to the backaction (thermal) heating.  

Our focus is on the backaction part.  In the typical experimental regime of weak coupling and low mechanical damping, the oscillator is mainly influenced by resonant bath modes, i.e. $\w \approx \Wm$.  
Near this frequency, the backaction heating is determined by a squeezed version of the input noise in Eq.~(\ref{eq:Bogo_bath}):
\begin{eqnarray}\label{eq:Bogo_TMSI}
\bA_1(\w) &\propto &
e^{i \phi_s(\w)}\cosh r_s(\w) \tAin_1[\w]  \\
&&+ e^{i \phi_s(\w)} e^{i \theta_s(\w)}\sinh r_s(\w) \tAind_1[\w] ~. \nonumber
\end{eqnarray}
The effective squeezing parameters are $e^{i \phi_s(\w)}\cosh r_s(\w) \equiv \Chi_{\tm,1}(\w)/\sqrt{|\Chi_{\tm,1}(\w)|^2-|\Chi_{\tm,1^\dag}(\w)|^2}$, $e^{i \theta_s(\w)} \tanh r_s(\w) \equiv \Chi_{\tm,1^\dag}(\w)/\Chi_{\tm,1}(\w)$;
$\bm{\Chi}(\w)\equiv i (\w\hI_6 - \bm{\mathcal{H}})^{-1}\bm{K}$ is the susceptibility matrix of the dressed modes; $\hI_k$ is the $k\times k$ identity.  

This picture explains that standard optomechanical cooling suffers from backaction heating because the mechanical oscillator is experiencing a photonic bath that is squeezed by the unwanted TMSI.  Our strategy is then to 
tune the PD so that the $\tAin_1$ in Eq.~(\ref{eq:Bogo_bath}) is \textit{already} squeezed vacuum noise, such that the additional squeezing in Eq.~(\ref{eq:Bogo_TMSI}) results in a simple vacuum noise in the vicinity of $\w \approx \Wm$.
This requires tuning
$2\phi_1+ \theta_1 = \theta_s(\Wm)$, and $r_1=r_s(\Wm)$, so the dominant bath mode reduces to vacuum noise:
\begin{equation}\label{eq:cool_require}
\bA_1(\Wm) \propto \Ain_1[\Wm]~;
\end{equation}
this can be satisfied if the parameters follow \cite{seeSI}
\begin{equation}\label{eq:cool_PD}
\lambda_1 = \Delta_1 - \Wm - i \kappa_1/2~.
\end{equation}
We note that although previous work provided numerical evidence that parametric driving could be beneficial  \cite{2009PhRvA..79a3821H, 2018PhyE..102...83A}, our analysis provides a physically-transparent and rigorous understanding of TMSI-induced backaction.  This understanding allows us to optimize the system parameters in Eq.~(\ref{eq:cool_PD}), which completely eliminates backaction heating and thus allows cavity cooling to the ground state in the deep sideband-unresolved regime.

\begin{figure}
\begin{center}
\includegraphics[width=\linewidth]{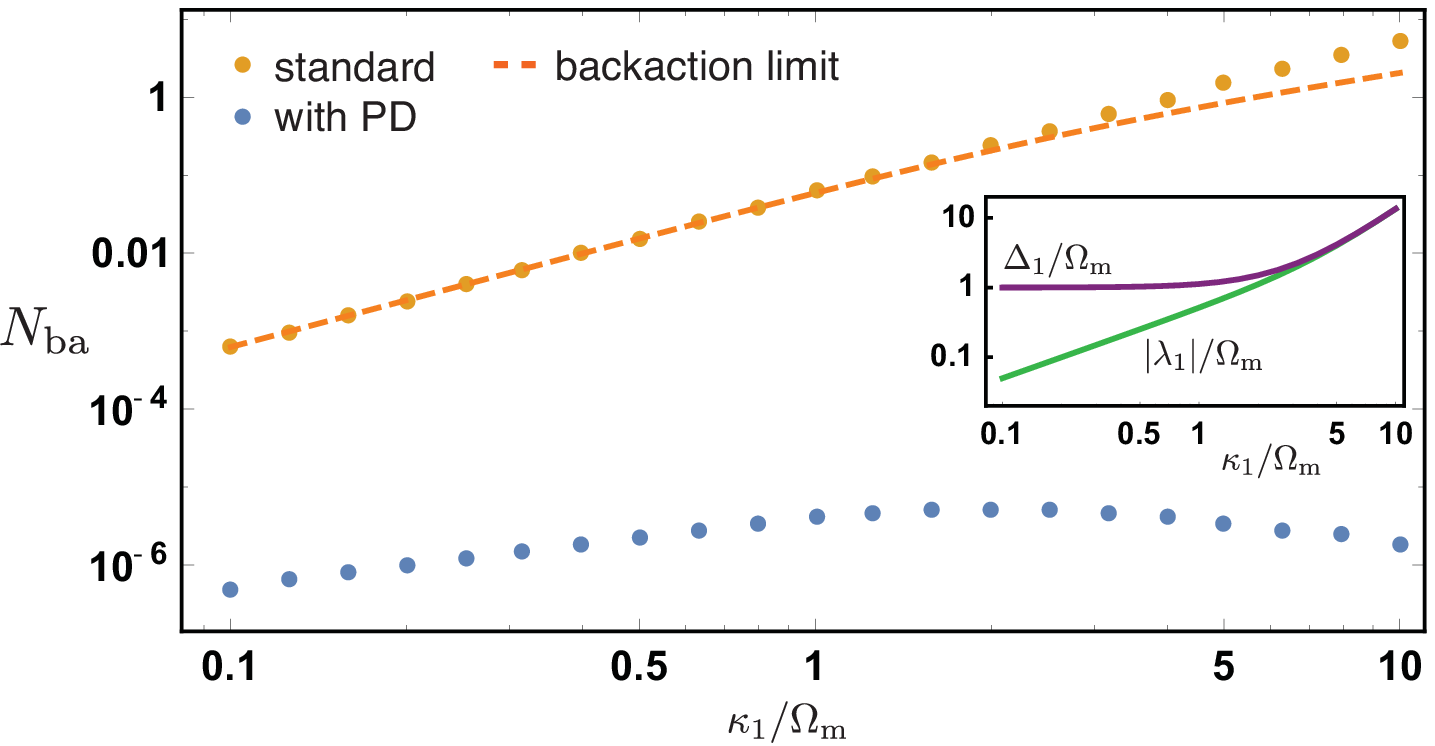}
\caption{Cavity backaction heating of the mechanics as a number of quanta $N_\textrm{ba}=\frac{1}{2\pi}\int \langle\bA_1^\dag(\w)\bA_1(\w') \rangle d\w d\w'$, versus sideband resolution parameter $\kappa_1/\Wm$.  $\tD = \Wm$, $\gamma=10^{-5}\Wm$, and $\sG_1$ is tuned to keep the cooperativity as $\sG_1^2/\kappa_1\gamma=10$.  At each value of $\kappa_1$, the standard (orange) and PD systems (blue) share the same $\bm{\mathcal{H}}$.  Optimal PD reduces $N_\textrm{ba}$ to below quantum backaction limit (dashed) $(\sqrt{1+(\kappa_1/2\Wm)^2}-1)/2$ \cite{WilsonRae:2007jp, Marquardt:2007dn}.  
The remaining excitation is due to non-vanishing but narrow mechanical linewidth ($\sim \sG_1^2/\kappa_1$) \cite{seeSI}.
(Inset)  Cavity mode detuning $\Delta_1$ and PD strength $|\lambda_1|$ for producing the same $\bm{\mathcal{H}}$.
\label{fig:Cooling}}
\end{center}
\end{figure}

Typical performance of our strategy is shown in Fig.~\ref{fig:Cooling}a, where the backaction excitation, $N_\textrm{ba}$, is suppressed far below the quantum backaction limit in both the sideband resolved ($\kappa_1 \ll \Wm$) and unresolved ($\kappa_1 \gtrsim \Wm$) regimes.  
We note that our PD strategy does not affect the heating of the mechanics by its intrinsic bath, nor requires increasing optomechanical coupling \cite{seeSI}.


Finally, our theory also provides a simple explanation of the injected squeezing (IS) strategy for sideband-unresolved cooling \cite{2017Natur.541..191C, 2016PhRvA..94e1801A}.  At first glance, both strategies are not obviously related: while the IS strategy requires arbitrarily strong squeezing in the bad cavity limit, the stationary intra-cavity squeezing generated in our PD approach is bounded by 3 dB \cite{MILBURN:1981tu}.  Despite this crucial difference, these strategies are connected: while we use PD to counteract the bath squeezing produced by TMSI, the IS strategy simply injects appropriately squeezed noise into the cavity to get the same kind of cancellation.  Explicitly, the goal is $\langle \bA^\dag_1(\w) \bA_1(\Wm)\rangle=0$, but now in Eq.~(\ref{eq:Bogo_TMSI}),  $\tAin_1[\w]$ represents vacuum noise, and $r_s(\w)$ and $\theta_s(\w)$ characterize the externally-produced squeezing.  Using this to determine optimal values of the squeezing parameters reduces to the same conditions found (via a slightly different argument) in Ref.~\cite{2017Natur.541..191C} \cite{seeSI}.  When comparing with the IS strategy, our approach has a crucial practical advantage:  it does not require one to externally produce and then transfer with high-fidelity a highly squeezed vacuum state.  This avoids the extra thermal noise induced due to transfer loss, which is a major limitating factor of the IS strategy  \cite{2017Natur.541..191C, 2016PhRvA..94e1801A}.  Nevertheless, these two strategies are complementary:  
the PD strength needed for perfect backaction suppression can be reduced if the photonic input noise is weakly squeezed.  

\textit{Quantum transduction--}  
Optomechanical quantum transduction is of enormous interest  
\cite{2011NJPh...13a3017S, 2011JPhCS.264a2025R, 2014NatPh..10..321A, Higginbotham:2018ca}.  
It requires coupling two cavities (one microwave, one optical) to a common mechanical oscillator, i.e.~our setup described in Eq.~(\ref{eq:LE_lab}).  We take the photonic baths to correspond to coupling waveguides, and take system 1 (2) to be the transducer input (output).  The transduction is characterized by the scattering of frequency modes $\vec{A}^\textrm{out}[\w]= \vec{A}^\textrm{in}[\w]+\bm{K} \vec{a}[\w]= \bm{T}(\w)\vec{A}^\textrm{in}[\w]$, where  $\bm{T}(\w) \equiv \hI_6 + i \bm{K}(\w\hI_6 - \bm{H})^{-1}\bm{K}$
is the scattering matrix.
An ideal transducer requires a frequency mode to be completely transferred,
i.e. $\Aout_2[\w_0] = \Ain_1[\w_0]$ at an optimal frequency $\w_0$.
In practice, such condition is usually not satisfied at any $\w$ due to system imperfections.  

To focus on the imperfection due to TMSI, we neglect the mechanical loss (i.e. $\gamma \rightarrow 0$).  The general transformation of a frequency mode is
\begin{equation}\label{eq:trans13}
\Aout_2[\w]= T_{2,1}(\w)\Ain_1[\w] + \hJ(\w)~,
\end{equation}
where  $\hJ(\w)\equiv \hJ_\textrm{T}(\w) + \hJ_\textrm{R}(\w)$ contains all the unwanted components being mixed into the transmitted mode by TMSI: $\hJ_\textrm{T}(\w)  \equiv T_{2,1^\dag}(\w)\Aind_1[\w]$ is introduced by the unwanted squeezing of the input, and the unwanted reflection is represented by
$\hJ_\textrm{R}(\w)\equiv T_{2,2}(\w) \Ain_2[\w]+T_{2,2^\dag}(\w) \Aind_2[\w]$.

In analogy to a linear amplifier, the performance of a bosonic transducer can be quantified by its added noise spectral density, i.e. how much extraneous noise is added to the output state
\cite{1982PhRvD..26.1817C}
\begin{equation}\label{eq:S_def}
2\pi \eta(\w) S (\w)\delta(\w -\w') \equiv \frac{1}{2} \left\langle \{ \hJ(\w), \hJ^\dag(\w')\} \right\rangle ~,
\end{equation} 
where $\eta(\w) \equiv |T_{2,1}(\w)|^2$ is the conversion efficiency.  For uncorrelated baths $1$ and $2$, their bosonic properties set a fundamental lower-bound on the added noise \cite{seeSI}:
\begin{equation}\label{eq:min_S}
S(\w)\geq \frac{\sR(\w)}{2} +\left|\frac{1-\eta(\w)}{2\eta(\w)} + \frac{\sR(\w)}{2} \right|~.
\end{equation}
Noiseless transduction thus requires a unit conversion efficiency, $\eta(\w) \rightarrow 1$, and a vanishing conjugated transmission (i.e. amplification), $\sR(\w)\equiv |T_{2,1^\dag}(\w)|^2/|T_{2,1}(\w)|^2 \rightarrow 0$, at an optimal frequency $\w \rightarrow \w_0$.

These conditions are generally not satisfied at any $\w$ for sideband unresolved optomechanical transducers, however 
we here show that they can be systematically achieved by parametrically driving only the input cavity $\ah_1$, and injecting squeezing to (but not parametrically driving) the output cavity $\ah_2$.  Our strategy again consists of tuning the PD and system parameters in such a way that the dressed-mode dynamical matrix $\bm{\mathcal{H}}$ (and hence dressed-mode scattering matrix $\bm{\sT}(\w) \equiv \hI_6 +\bm{K}\bm{\Chi}(\w)$) remains unchanged.  
The relation between the dressed-mode and original-mode scattering matrices is:
\begin{equation}\label{eq:T_lab_eff}
\vec{A}^\textrm{out}[\w] =\bm{T}(\w) \vec{A}^\textrm{in}[\w] = \sF^{-1} \bm{\sT}(\w) \sF \vec{A}^\textrm{in}[\w]~,
\end{equation}
where the PD-induced squeezing is described by 
 $\sF_1 \equiv 
\left(\begin{array}{cc} e^{i\phi_i}\cosh r_1 & -e^{i\phi_1} e^{i\theta_1}\sinh r_1 \\ -e^{-i\phi_1} e^{-i\theta_1}\sinh r_1 & e^{-i\phi_1}\cosh r_1 \end{array}\right) $ and $\sF \equiv \textrm{diag}(\sF_1, \hI_4)$.
For simplicity, we assume all dressed modes are resonant, i.e. $\tD_1=\tD_2=\Wm$, although a generalization beyond this regime is straightforward.

To correct the unwanted amplification (i.e.~$\mathcal{R}(\omega) \neq 0$), 
we first consider the transmission block of the scattering matrix in Eq.~(\ref{eq:T_lab_eff}), which gives
$T_{2,1^\dag}(\w) = e^{-i\phi_1}\cosh r_1 \sT_{2,1^\dag}(\w)-e^{i\phi_1} e^{i\theta_1}\sinh r_1 \sT_{2,1}(\w) $.
At any specific $\w$, the TMSI-induced amplification can be corrected by a PD-induced squeezing, such that $T_{2,1^\dag}(\w)=0$.   
The required squeezing parameters follow
\begin{equation}\label{eq:trans_param}
e^{i2\phi_1} e^{i \theta_1} \tanh r_1 = \sT_{2,1^\dag}(\w)/\sT_{2,1}(\w)~.
\end{equation}

Apart from correcting unwanted amplification by PD, we also need to choose the system parameters that yield unity conversion efficiency $\eta(\omega) = 1$.  From the detailed expression of conversion efficiency \cite{seeSI}, we find that it is maximized when $\Gamma_1 = \Gamma_2$, where
\begin{equation}\label{eq:Gamma}
\Gamma_i \equiv  \frac{4 \sG_i^2}{\kappa_i} - \frac{(\kappa_i/4\Wm)^2}{1+(\kappa_i/4\Wm)^2}\frac{4 \sG_i^2}{\kappa_i}~.
\end{equation}
$\Gamma_i$ is nothing but the net optical damping rate of the mechanics due to cavity; the condition $\Gamma_1 = \Gamma_2$ can thus be seen as a generalized impedance matching condition.

Because of the TMSI-induced amplification, the peak efficiency will in general be higher than unity ($\max(\eta)>1$), which prevents optimizing the added noise.  To optimize this added noise over a reasonable bandwidth, it is thus desirable to {\it deliberately} impedance mismatch the system so that $\max(\eta) = 1$.  This requires satisfying the modified impedance matching condition \cite{seeSI}: 
\begin{equation}\label{eq:optimal_mismatch}
\left(1+ (\frac{\kappa_2}{4 \Wm})^2 \right)\left(\frac{\sqrt{\Gamma_1 \Gamma_2}}{(\Gamma_1 +\Gamma_2)/2} \right)^2=1~.
\end{equation}
With this choice, the conversion efficiency is close to unity for the frequency modes around $\w_0= \Wm - (\kappa_1 \Gamma_1 + \kappa_2 \Gamma_2)/8\Wm$.

Finally, achieving the lower bound in Eq.~(\ref{eq:min_S}) 
also requires optimizing the input noise injected into the {\it output} of our transducer (i.e.~$\hat{a}_2$).  
At the optimal frequency where $\eta(\w_0)=1$ and $\sR(\w_0)=0$ (hence $\hat{J}_\textrm{T}(\w_0)=0$), 
the added noise only involves the cavity-2 bath, via the operator $\hat{J}_\textrm{R}(\w_0)$.
We find that the commutation relation of Eq.~(\ref{eq:trans13}) imposes $|T_{2,2}(\w_0)|=|T_{2,2^\dag}(\w_0)|$, which requires the real and imaginary parts of $\hat{J}_\textrm{R}(\w_0)$ to be two distinct but {\it commuting} quadrature operators.  As such, the added noise can be suppressed by injecting to cavity-2 a bath that is squeezed in both of these quadratures.  
In practice, this squeezing is achievable by having broadband single-mode squeezing.
Explicitly, we consider a squeezed bath with correlations $\langle \Aind_2(t)\Ain_2(t')\rangle = \delta(t-t') \sinh^2 s$ and $\langle \Ain_2(t)\Ain_2(t')\rangle = \delta(t-t')e^{i\vartheta}\sinh s \cosh s $.  By evaluating Eq.~(\ref{eq:S_def}), the added noise at $\w=\w_0$ is suppressed for increasing squeezing strength $s$:
$S(\w_0) = e^{-2s}|T_{2,2^\dag}(\w_0)|^2$ when the squeezing phase is optimized as
$e^{i\vartheta} = -T_{2,2^\dag}(\w_0)/T_{2,2}(\w_0)$.

\begin{figure}
\begin{center}
\includegraphics[width=\linewidth]{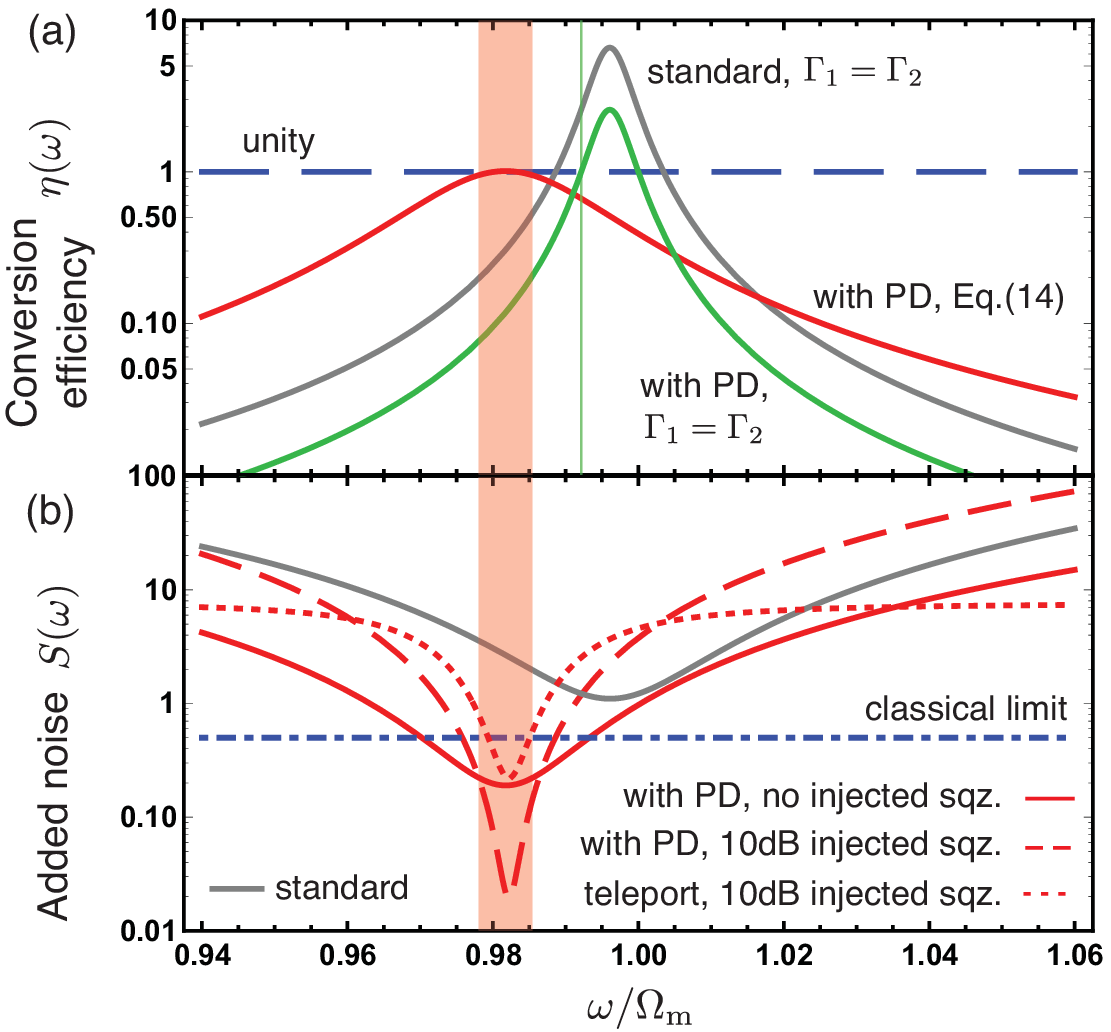}
\caption{(a) Conversion efficiency $\eta(\w)$ of optomechanical transducers with $\tD_1=\tD_2=\Wm$, $\kappa_1=\kappa_2=5\Wm$, and $\sG_2= 0.1\Wm$.  
(Grey) Standard transducer: no PD and 
$\Gamma_1=\Gamma_2$.
(Green) Optimal PD in (\ref{eq:trans_param}) with $\Gamma_1=\Gamma_2$.
(Red) Our strategy: optimal PD and modified impedance matching in Eq.~(\ref{eq:optimal_mismatch}).  
Transduction windows with $|1-\eta(\w)| < 5 \%$ are shown for the PD systems.  Modified impedance matching (red shaded) yields a much wider bandwidth than picking $\Gamma_1=\Gamma_2$ (green shaded).
(b) Added noise $S(\w)$ for the transducers.
(Dot-dashed) $S(\w)=0.5$, above which more noise is added in coherent state transduction than classical cloning \cite{Braunstein:2005wr, Transduction_interference}.  
(Grey) Standard transducer adds more noise than the classical limit in this scenario, i.e. no quantum transduction.
Our strategy can yield a transduction with lower noise even when the bath is vacuum (red solid); the noise can further be suppressed by injecting 10 dB squeezing (red dashed).
(Red dotted) Teleportation strategy with 10 dB input two-mode-squeezing and 10 dB injected squeezing at output port.  
\label{fig:Transduction}}
\end{center}
\end{figure}

The performance of a typical optomechanical transducer that is only weakly sideband resolved is shown in Fig.~\ref{fig:Transduction}.  As shown, our approach gives a marked improvement in the transduction fidelity.    We note that even when there is mechanical loss, $\gamma \neq 0$, our PD strategy would not amplify the additional contribution to the transducer added noise coming from the intrinsic mechanical bath.
This is because such noise is determined by the scattering amplitudes $T_{2,m}(\w)$ and $T_{2,m^\dag}(\w)$, which is unaltered if the system parameters are tuned to preserve  $\bm{\mathcal{H}}$ (c.f. Eq.~(\ref{eq:T_lab_eff})).

Our strategy can improve the performance of microwave-optics transducer, which is a crucial component in a superconducting quantum computer network \cite{2011NJPh...13a3017S}.  For microwave-to-optics transfer, our strategy requires a microwave-cavity PD and injected optical squeezing, which both have been realized with high quality \cite{2008ApPhL..93d2510Y, 2008NatPh...4..929C, 2011PhRvL.106k3901F, 2016PhRvL.117k0801V}.  For the reverse direction (i.e.~optics-to-microwave), an optical-cavity PD is instead required.  With numerous exciting prospects  \cite{2015PhRvL.115x3603P,2015PhRvL.114i3602L, Leroux:2018kj, 2018PhRvL.120i3601Q}, optical-cavity PD has been experimentally implemented by embedding nonlinear crystals in optical cavities \cite{2008PhRvL.100c3602V, 2016PhRvL.117k0801V, 2017PhR...684....1S} or fabricating microcavities with nonlinear materials \cite{2011PhRvL.106k3901F, 2013NatCo...4.1818F}.

Alternatively, one can apply our strategy for optics-to-microwave transfer {\it without} any need for an optical-cavity PD by combining it with quantum teleportation \cite{Braunstein:2005wr, 2015NaPho...9..641P} (more details and schematic illustration are given in Supplementary Material).  One first prepares a microwave two-mode-squeezed state, injecting one half into a microwave-to-optics transducer, emerging at the output of the optical cavity.  The input optical state to be transduced is not injected into the transducer; instead, one makes a continuous-variable Bell measurement on this state and the optical-cavity output state.  After feedforward, the original optical input state can be recovered from the remaining half of the microwave two-mode-squeezed state.  This method replaces the technically challenging optical parametric drive by highly efficient microwave squeezing \cite{2012PhRvL.109y0502M, 2014PhRvL.113k0502E, 2016PhRvL.117b0502F, 2017PhRvX...7d1011B} and optical homodyne detection \cite{2015NatCo...6E6665F}.

\textit{Conclusion--}  In this work we study the backaction effects of TMSI in sideband unresolved optomechanical systems.  We show that most detrimental effects originate from the squeezing of photonic bath, which can be corrected by a controlled squeezing through parametrically driving the photonic cavity.  We show explicitly how our strategy can eliminate backaction heating in optomechanical cooling, and correct unwanted amplification in quantum transduction.  
Although our analysis is focused on optomechanical systems, the technique is also applicable to eliminate TMSI-induced unwanted effects in other hybrid quantum platforms, such as electro-optical \cite{2010PhRvA..81f3837T, 2011PhRvA..84d3845T, Rueda:2016jg} and magnomechanical systems \cite{2016SciA....2E1286Z}.

\textit{Acknowledgement--}  This work is supported by the AFOSR MURI FA9550-15-1-0029 on quantum transduction.

\textit{Note added--}  During the review process of this work, we became aware of the appearance of related articles, Refs.~\cite{Asjad:2019, Gan:2019ef}, which also discuss improved optomechanical cooling with parametric drive.  However, these works do not consider the application of parametric drive in quantum transduction.

\onecolumngrid
\section{Supplementary Material}

\subsection{Stability of optomechanics systems}

We here provide a brief discussion about the stability of our optomechanical systems.  By construction, we consider a strength of parametric drive that is bounded by the mode detuning, i.e. $|\lambda_i| < |\Delta_i|$.  By considering the dynamical matrix of $\ah_i$ and $\ad_i$, it is straightforward to show that this condition is sufficient for the cavity modes to be stable in the absence of optomechanical coupling.
Such stability guarantees that the Hamiltonian of each cavity can be diagonalized by Bogoliubov transformation, i.e. the transformation parameters in Eq.~(\ref{eq:Bogo_para}) are physical.  Under a valid Bogoliubov transformation, the stability of the parametrically driven system in Eq.~(\ref{eq:LE_lab}) is equivalent to that of the dressed mode dynamics in Eq.~(\ref{eq:LE_eff}).  It is because the stability of a physical system is not affected by whether the system dynamics is described by cavity modes or dressed modes.

As a linear system, the stability of Eq.~(\ref{eq:LE_eff}) is solely determined by its dynamical matrix $\bm{\mathcal{H}}$.  Generally, this matrix also describes the dynamics of a standard (i.e. without parametric drive) optomechanical system with the modified mode detunings $\tD_i$ and optomechanical couplings $\sG_i$.  While the stability analysis for standard optomechanical systems is outside the scope of this paper, for the aim of our strategy we focus on the systems with red detuning, i.e. $\tD_i > 0$.  We note that a particular regime of this problem was considered in Ref.~\cite{2015PhRvA..91a3807W}.

For optomechanical cooling, which involves one cavity and one oscillator, we verify the system stability by checking the Routh-Hurwitz criterion of the eigenvalue equation $\det(-i \bm{\mathcal{H}}-\sigma \hI) =0$ \cite{1987PhRvA..35.5288D}, where $\sigma$ is the eigenvalue.  In this case, the generalized Sturm chain has the sign \cite{1987PhRvA..35.5288D}
\begin{eqnarray}
P_0(\infty)>0 ~~, ~~ P_1(\infty)<0 ~~,~~ P_2(\infty)>0 ~~,~~ P_3(\infty)<0 \nonumber \\
P_0(-\infty)>0 ~~, ~~ P_1(-\infty)>0 ~~,~~ P_2(-\infty)>0 ~~,~~ P_3(-\infty)>0~.
\end{eqnarray}
The system stability is solely determined by the sign of the last coefficient in the generalized Sturm chain, $P_4$.  From the exact expression of $P_4$, we find that the system is unstable if and only if
\begin{equation}
\sG_1 > \sqrt{\frac{1}{4 \tD_1 \Wm} (\tD_1^2 +\frac{\kappa_1^2}{4}) (\Wm^2+\frac{\gamma^2}{4}) } 
\end{equation}
This condition implies that the system is stable unless the optomechanical coupling is in the ultra-strong regime \cite{Peterson:2019}, which is way beyond the weak coupling regime that we are interested in.

For optomechanical transducer, which involves two cavities and one mechanical oscillator, analytically verifying the Routh-Hurwitz criterion is unfortunately difficult due to the abundance of system parameters.  From the sign of the last coefficient in the generalized Sturm chain, we can find only a sufficient condition of instability when
\begin{equation}
\frac{\sG_1^2 \tD_1}{\tD_1^2 + \kappa_1^2/4} + \frac{\sG_2^2 \tD_2}{\tD_2^2 + \kappa_1^2/4} 
> \frac{1}{4} \frac{\Wm^2 + \gamma^2/4}{\Wm}  
\end{equation}
which can be satisfied only in the ultra-strong coupling regime.  Nevertheless, we are mainly interested in the weak coupling regime which is implemented in most experiments, i.e. $\sG_1, \sG_2 \ll \tD_i, \kappa_i $.  In this regime, we can treat the optomechanical coupling as perturbation, and verify the system stability by computing the leading order eigenvalue perturbation.  For the photonic modes, the optomechanics-induced gain and loss rate is perturbative when comparing to their bare dissipation, $\kappa_i$, and so the photonic modes are stable.  For the mechanical oscillator, however, its dissipation rate is assumed weak in order to yield a high cooperativity, i.e. $\gamma \ll \sG_i$.  At the leading order, we find that the optomechanical couplings induce a dissipation on the mechanical oscillator, i.e.
\begin{equation}
\textrm{Re}(\sigma_\textrm{m}) \approx -\frac{\gamma}{2} 
- \frac{2 \sG_1^2 \kappa_1 \tD_1 \Wm}{\big((\tD_1-\Wm)^2+\kappa_1^2/4 \big) \big((\tD_1+\Wm)^2+\kappa_1^2/4 \big)}
- \frac{2 \sG_2^2 \kappa_2 \tD_2 \Wm}{\big((\tD_2-\Wm)^2+\kappa_2^2/4 \big) \big((\tD_2+\Wm)^2+\kappa_2^2/4 \big)}~.
\end{equation}
Hence both the cavity modes and mechanical oscillator are stable in the weak coupling regime.  We note that all the systems simulated in this work are numerically verified to be stable.


We emphasize that when applying our strategy, the first step is to construct a stable dressed mode system (i.e. a stable $\bm{\mathcal{H}}$).  After that, the corresponding system parameters for a parametrically driven system (i.e. $\bm{H}$) are inversely deduced from Eq.~(\ref{eq:Bogo_para}) as well as the relations between $\tD_i, \sG_i$ and $\Delta_i, G_i$

\subsection{Reproducing results in Ref.~\cite{2017Natur.541..191C}}

We here provide a simple derivation for the optimal injected squeezing in Ref.~\cite{2017Natur.541..191C} that eliminates backaction heating.  Firstly, because the cavity is not parametrically driven, the input field is only squeezed by TMSI, i.e. for the frequency modes near resonance, $\w\approx \Wm$, the mechanical oscillator response to photonic bath is
\begin{equation}
\bA(\w)= \mathbb{X}(\w) \bbA(\w),~~\textrm{where~}
\bbA(\w)\equiv \frac{\chi_{m,1}(\w)\Ain_1[\w] + \chi_{m,1^\dag}(\w)\Aind_1[\w] }{\sqrt{|\chi_{m,1}(\w)|^2-|\chi_{m,1^\dag}(\w)|^2}} 
\end{equation}
and $\mathbb{X}(\w) = \sqrt{|\chi_{m,1}(\w)|^2-|\chi_{m,1^\dag}(\w)|^2}$.  $\bbA(\w)$ can be seen as a squeezed noise operator because it obeys the canonical commutation relation, $[\bbA(\w),\bbA^\dag(\w')]=2\pi \delta(\w-\w')$.  

In the weak coupling regime, the backaction excitation is mainly determined by the mean photon occupation at the resonant frequency, i.e. $\langle \bA^\dag(\w) \bbA(\Wm) \rangle=n(\Wm) 2 \pi \delta(\w-\Wm)$.  For an injected squeezing bath with $\langle \Ain_1(t)\Ain_1(t')\rangle = \delta(t-t') e^{i\theta_b} \sinh r_b \cosh r_b$ and $\langle \Aind_1(t)\Ain_1(t')\rangle = \delta(t-t') \sinh^2 r_b$, the mean photon occupation is
\begin{equation}
n(\Wm) = \frac{| \chi_{m,1}(\Wm) e^{i\theta_b} \sinh r_b +  \chi_{m,1^\dag}(\Wm) \cosh r_b|^2}{| \chi_{m,1}(\Wm)|^2 -| \chi_{m,1^\dag}(\Wm)|^2}~.
\end{equation}
As a result, the optimal squeezing that yields $n(\Wm) =0$ obeys
\begin{equation}\label{eq:ext_sq}
e^{i \theta_b} \tanh r_b = -\frac{\chi_{m,1^\dag}(\Wm)}{\chi_{m,1}(\Wm)}= \frac{\Wm -\Delta_1  +i \kappa_1/2 }{\Wm+ \Delta_1  +i \kappa_1/2}~.
\end{equation}

After straightforward algebra, we have
\begin{equation}
\cosh^2 2r_b = \frac{1}{(2\Wm)^2} \left(\Delta_1+ \frac{\Wm^2 + (\kappa_1/2)^2}{\Delta_1} \right)^2 
\geq 1+ \left(\frac{\kappa_1}{2 \Wm} \right)^2~,
\end{equation}
where the last bound is attainable when the cavity detuning is chosen as $\Delta_1 = \sqrt{\Wm^2+(\kappa_1/2)^2}$.  This bound places the minimum squeezing required to completely suppress the backaction heating.  Rearranging the expression, we reproduce Eq.~(5) in Ref.~\cite{2017Natur.541..191C}:
\begin{equation}
\sinh 2r_b \geq \frac{\kappa_1}{2 \Wm}~.
\end{equation}  

Apart from the squeezing strength, the phase of squeezing can be obtained from Eq.~(\ref{eq:ext_sq}) as
\begin{equation}
\tan \theta_b = \frac{\Delta_1 \kappa_1}{\Wm^2 - \Delta_1^2 +(\kappa_1/2)^2}~,
\end{equation}
which is the same as Eq.~(11) in Ref.~\cite{2017Natur.541..191C} (note that the detuning in Ref.~\cite{2017Natur.541..191C} is equivalent to $-\Delta_1$ in this work).

\subsection{Derivations of the minimum added noise in Eq.~(\ref{eq:min_S})}

Consider the R.H.S. of Eq.~(\ref{eq:S_def}):
\begin{equation}
\frac{1}{2} \left\langle \{ \hJ(\w), \hJ^\dag(\w')\} \right\rangle = \frac{1}{2} \left\langle \{ \hJ_\textrm{T}(\w), \hJ_\textrm{T}^\dag(\w')\} \right\rangle  + \frac{1}{2} \left\langle \{ \hJ_\textrm{R}(\w), \hJ_\textrm{R}^\dag(\w')\} \right\rangle~,
\end{equation}
where the input noises to cavities 1 and 2 are assumed to be uncorrelated and zero mean.  The first term denotes the noise caused by the TMSI-induced squeezing of the transmitted signal, 
\begin{eqnarray}
\frac{1}{2} \left\langle \{ \hJ_\textrm{T}(\w), \hJ_\textrm{T}^\dag(\w')\} \right\rangle = \frac{1}{2}\left\langle[ \hJ_\textrm{T}^\dag(\w), \hJ_\textrm{T}(\w')]  \right\rangle + \left\langle \hJ_\textrm{T}(\w) \hJ_\textrm{T}^\dag(\w') \right\rangle 
&=& \frac{1}{2} |T_{2,1^\dag}(\omega)|^2 2 \pi \delta(\w - \w') + \left\langle \hJ_\textrm{T}(\w) \hJ_\textrm{T}^\dag(\w') \right\rangle \nonumber \\
&\geq&  \frac{1}{2} |T_{2,1^\dag}(\omega)|^2 2 \pi \delta(\w - \w')~.
\end{eqnarray}
The last bound can be satisfied in the typical situation, where the input noise to cavity 1 is vacuum (except the frequency mode that contains the quantum information).

The second term denotes the noise caused by the non-vanishing reflection of input noise at cavity 2.  Consider the relation
\begin{eqnarray}
\frac{1}{2} \left\langle \{ \hJ_\textrm{R}(\w), \hJ_\textrm{R}^\dag(\w')\} \right\rangle &=& \frac{1}{2}\left\langle[ \hJ_\textrm{R}^\dag(\w), \hJ_\textrm{R}(\w')]  \right\rangle + \left\langle \hJ_\textrm{R}(\w) \hJ_\textrm{R}^\dag(\w') \right\rangle = \frac{1}{2}\left\langle[ \hJ_\textrm{R}(\w), \hJ_\textrm{R}^\dag(\w')]  \right\rangle + \left\langle \hJ_\textrm{R}^\dag(\w) \hJ_\textrm{R}(\w') \right\rangle
 \nonumber \\
&\geq&  \frac{1}{2} \left| [ \hJ_\textrm{R}^\dag(\w), \hJ_\textrm{R}(\w')]  \right|  \nonumber \\
&=& \frac{1}{2} \left| |T_{2,2}(\w)|^2 - |T_{2,2^\dag}(\w)|^2 \right| 2 \pi \delta(\w-\w')~.
\end{eqnarray}
The lower bound at the second line is attainable when the input noise produces a vanishing expectation value $\left\langle \hJ_\textrm{R}^\dag(\w) \hJ_\textrm{R}(\w') \right\rangle$ or $\left\langle \hJ_\textrm{R}(\w) \hJ_\textrm{R}^\dag(\w') \right\rangle$.  Depending on the magnitude of $|T_{2,2}(\w)|$ and $|T_{2,2^\dag}(\w)|$, either $\hJ_\textrm{R}(\w)$ or $\hJ_\textrm{R}^\dag(\w)$ represents a squeezed noise operator.  The lower bound is thus attainable by injecting squeezed noise to cavity 2.

Due to the preservation of commutation relation, Eq.~(\ref{eq:trans13}) implies that the transformation amplitudes obey
\begin{equation}
1= |T_{2,1}(\w)|^2 - |T_{2,1^\dag}(\w)|^2 +  |T_{2,2}(\w)|^2 - |T_{2,2^\dag}(\w)|^2~.
\end{equation}
Substituting into this relation with the definitions of conversion efficiency, conjugate transmission, and added noise in Eq.~(\ref{eq:S_def}), the lower bound of added noise in Eq.~(\ref{eq:min_S}) can be obtained.

\subsection{Mechanical susceptibility modified by parametric drive}

\begin{figure}
\begin{center}
\includegraphics[width=0.7\linewidth]{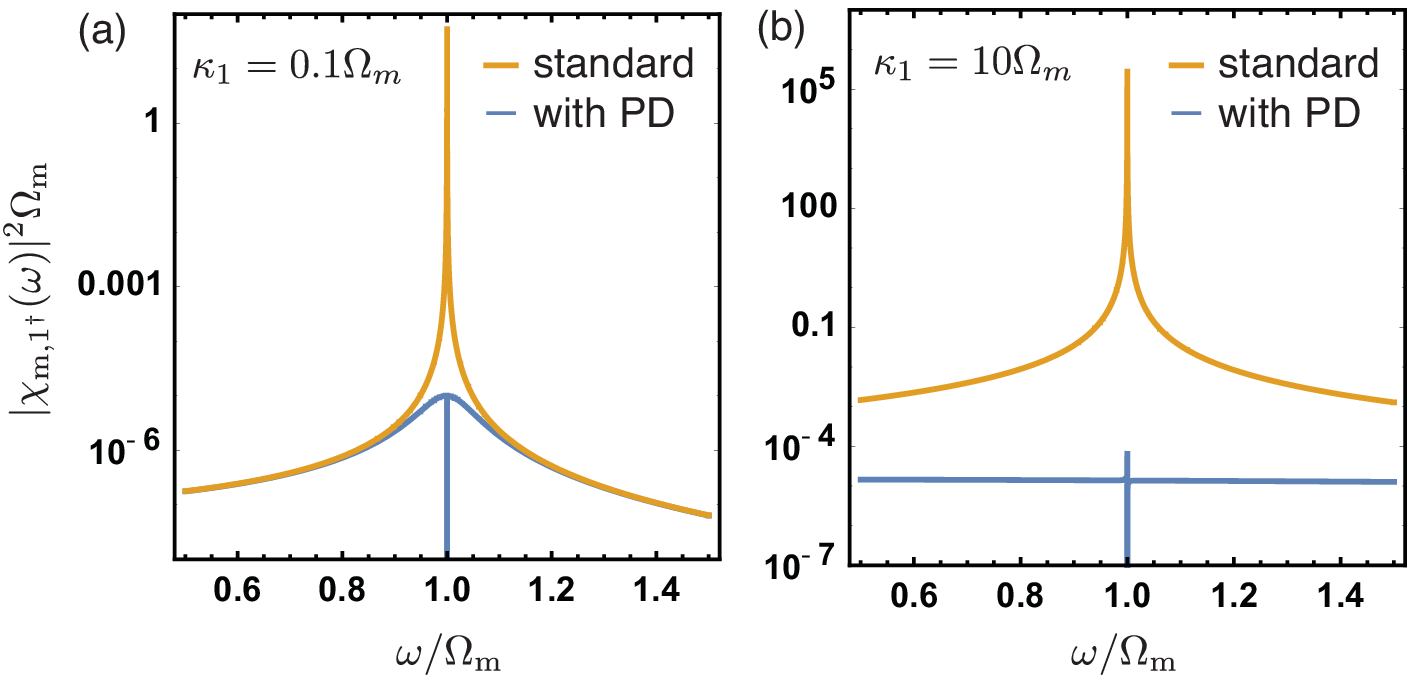}
\caption{ In both sideband resolved (a) and unresolved regimes (b), optimal PD significantly suppresses the spectrum $|\chi_{\tm,1^\dag}(\w)|^2$ at $\w\approx \Wm$, which contributes dominantly to $N_\textrm{ba}$ in Eq.~(\ref{eq:Nba}).
\label{fig:Cooling_b}}
\end{center}
\end{figure}

While in the dressed mode picture, our strategy can be understood as a compensating squeezing transformation, we stress that we are not changing the form of the input fluctuations in the lab frame of Eq.~(\ref{eq:LE_lab}).  As such, our results can also be interpreted as the PD modifying the appropriate susceptibility of the mechanics.  The susceptibility matrix of our system (in terms of the original photon modes) is 
$\bm{\chi}(\w)\equiv i (\w\hI_6 - \bm{H})^{-1}\bm{K}$.  The backaction excitation can then be written as
\begin{equation}\label{eq:Nba}
	N_\textrm{ba}=\frac{1}{2\pi}\int \langle\bA_1^\dag(\w)\bA_1(\w') \rangle d\w d\w'=\frac{1}{2\pi}\int |\chi_{m,1^\dag}(\w)|^2 d\w~,
\end{equation}
We plot in Fig.~\ref{fig:Cooling_b} the effective excitation spectrum $|\chi_{m,1^\dag}(\w)|^2$.  One sees that our strategy uses PD and parameter tuning to suppress the spectrum at the dominant frequencies $\w\approx \Wm$.

\subsection{Optimal parametric drive for vanishing backaction heating, Eq.~(\ref{eq:cool_PD}) }

To verify optimal PD strength and detuning in Eq.~(\ref{eq:cool_PD}), we can compute the squeezing parameters as in Eqs.~(\ref{eq:Bogo_para}) and (\ref{eq:Bogo_TMSI}), and relate them according to the requirement in Eq.~(\ref{eq:cool_require}).  Such procedure is straightforward but tedious.  

Alternatively, we note that our results can also be interpreted as an optimal PD-induced modification of mechanical susceptibility, apart from being understood as a squeezing compensation.
Therefore, we can verify Eq.~(\ref{eq:cool_PD}) by considering the relation
\begin{equation}
\bA_1(\w) = \Chi_{m,1}(\w) \tAin_1[\w] + \Chi_{m,1^\dag}(\w) \tAind_1[\w] =  \chi_{m,1}(\w) \Ain_1[\w] + \chi_{m,1^\dag}(\w) \Aind_1[\w]~.
\end{equation}
The condition in Eq.~(\ref{eq:cool_require}) is then equivalent to having $\chi_{m,1^\dag}(\Wm)=0$.  Consider the explicit form of this susceptibility,
\begin{equation}
\chi_{m,1^\dag}(\w) = \sqrt{\kappa_1}\frac{G_1 \left(\lambda_1 + (\w- \Delta_1) +i\frac{\kappa_1}{2} \right)\left(\w + \Wm +i \frac{\gamma}{2} \right)}
{\left( (\w +i \frac{\kappa_1}{2})^2-\Delta_1^2 +|\lambda_1|^2 \right)((\w+i\frac{\gamma}{2})^2-\Wm^2)- 4 G_1^2 \Delta_1 \Wm + 2G_1^2 \Wm (\lambda_1 + \lambda_1^\ast)} ~.
\end{equation}
From the numerator, it is then obvious that Eq.~(\ref{eq:cool_PD}) produces a vanishing $\chi_{m,1^\dag}(\Wm)$.  

We note that when the optomechanical coupling is not infinitesimally weak, our strategy leaves a small residual backaction heating.  The reason is easily understood: the TMSI-induced squeezing is frequency dependent, whereas the PD-induced squeezing is constant as a function of frequency.  This makes perfect cancellation at all frequencies impossible.  As long as $\sG_1 < \kappa_1$, this imperfection is however extremely weak, as the total mechanical linewidth ($\sim \sG_1^2/\kappa_1$) is much small than the frequency range ($\sim\kappa_1$) over which the TMSI-induced squeezing varies.

\subsection{Conservation of mechanical heating and coupling}

Crucially, our strategy does not have any impact on the heating of the mechanics by its intrinsic bath.  This can be observed from the response of the oscillator to the mechanical bath, i.e. 
\begin{equation}
\bA_\textrm{m}(\w) \equiv \Chi_{m,m}(\w) \Ain_\textrm{m}[\w] + \Chi_{m,m^\dag}(\w) \Aind_\textrm{m}[\w]~.
\end{equation}
Because the system parameters are tuned to preserve the dressed-mode dynamics, $\bm{\mathcal{H}}$, both the susceptibilities $\Chi_{m,m}(\w)$ and $\Chi_{m,m^\dag}(\w)$, and hence $\bA_\textrm{m}(\w)$ are unaltered.
Therefore the cooling performance is unaffected by our strategy.
Generally, in order to preserve the dressed mode detuning $\tD_1$ as $\lambda_1$ is increased, the actual drive detuning $\Delta_1$ in Eq.~(\ref{eq:LE_lab}) also has to be increased (Fig.~\ref{fig:Cooling}a inset).
In addition, the the many-photon optomechanical coupling $G_1$ should be adjusted to keep the dressed mode coupling $\sG_1$ constant.
We stress that, however, the improvement provided by our strategy is not simply due to a stronger coupling \cite{2018PhyE..102...83A}.  In fact, with the parameters in Fig.~\ref{fig:Cooling} both the standard and PD systems (with the same $\kappa_1$) share the same strength of $G_1$.

\subsection{Maximum conversion efficiency and impedance matching }

We consider the conversion efficiency $\eta(\w)$ of a parametrically driven optomechanical transducer.  By choosing the squeezing parameters as in Eq.~(\ref{eq:trans_param}), the input amplification is cancelled and the conversion efficiency is given by
\begin{eqnarray}
\eta(\w)=|T_{2,1}(\w)|^2 = |\sT_{2,1}(\w)|^2 - |\sT_{2,1^\dag}(\w)|^2 =  \qquad \nonumber \\
 \left|\frac{4 \sG_1 \sG_2 ((\w+\Wm)+i\kappa_2/2) \sqrt{\kappa_1 \kappa_2}\Wm \w }{((\w+i\kappa_1/2)^2 - \Wm^2)((\w+i\kappa_2/2)^2 - \Wm^2)(\w^2 - \Wm^2) - 4\sG_1^2 \Wm^2 ((\w+i\kappa_2/2)^2 - \Wm^2) - 4\sG_2^2 \Wm^2 ((\w+i\kappa_1/2)^2 - \Wm^2) } \right|^2 ~.  \nonumber \\
\end{eqnarray}
In the weak coupling regime, i.e. $\sG_1, \sG_2 \ll \kappa_1,\kappa_2, \Wm$, around the resonant frequency $\w \approx \Wm$ the conversion efficiency can be approximated as a Lorentzian, i.e. 
\begin{eqnarray}
\eta(\w) &\approx  &
\left|\frac{4 \sG_1 \sG_2 (\Wm+i\kappa_2/4) \sqrt{\kappa_1 \kappa_2}\Wm }{((i\Wm \kappa_1- \kappa_1^2/4 )(i\Wm \kappa_2 - \kappa_2^2/4)\left(\w - \Wm +i \frac{4}{1+i\kappa_1/4\Wm} \frac{\sG_1^2}{2\kappa_1} +i \frac{4}{1+i\kappa_1/4\Wm} \frac{\sG_2^2}{2\kappa_2}\right) } \right|^2 \\
&=& \left(1+ (\frac{\kappa_2}{4 \Wm})^2 \right) \left|\frac{\sqrt{\Gamma_1\Gamma_2}}{\w -\w_0 -i (\Gamma_1+\Gamma_2)/2}  \right|^2~,  \label{eq:eta_full}
\end{eqnarray}
where $\w_0 = \Wm - (\kappa_1 \Gamma_1+ \kappa_2 \Gamma_2)/8\Wm$ and $\Gamma_i$ is defined in Eq.~(\ref{eq:Gamma}).

Eq.~(\ref{eq:eta_full}) is generic for any optomechanical transducer in the weak coupling regime.  It is obvious that the conversion efficiency is peaked for the frequency mode $\w=\w_0$, i.e.
\begin{equation}\label{eq:eta_w0}
\eta(\w) \leq \eta(\w_0) = \left(1+ (\frac{\kappa_2}{4 \Wm})^2 \right) \left(\frac{\sqrt{\Gamma_1\Gamma_2}}{(\Gamma_1+\Gamma_2)/2}  \right)^2~.
\end{equation}
In analogous to impedance matching in the resolved sideband regime, the conversion efficiency is maximum for a transducer with specific $\Wm$ and $\kappa_2$ when the optomechanical couplings ($\sG_1$, $\sG_2$) are tuned (by adjusting the cavity drive) to equate, i.e.
\begin{equation}
\textrm{when}~\Gamma_1=\Gamma_2,~~~\eta(\w_0) = 1+ \big( \frac{\kappa_2}{4\Wm}\big)^2 >1~.
\end{equation}
However, as discussed in the main text, for a wide bandwidth transduction the conversion efficiency is desirable to peak at $\eta(\w_0)=1$.  This condition requires $\Gamma_1 \neq \Gamma_2$, instead $\sG_1$ and $\sG_2$ (or equivalently $\Gamma_1$ and $\Gamma_2$) are tuned that Eq.~(\ref{eq:eta_w0}) is equal to unity.  This yields the modified impedance matching condition in Eq.~(\ref{eq:optimal_mismatch}).

Finally, although mechanical dissipation has been neglected in this work for simplicity, we have also computed but not shown the general scattering matrix with $\gamma\neq 0$.  We find that our analysis remains valid when $\Gamma_i \gg \gamma$.  This is the regime where the mechanical oscillator couples stronger to the input and output photonic baths than to the mechanical bath.

\subsection{Teleportation-based strategy}

\begin{figure}
\begin{center}
\includegraphics[width=0.9\linewidth]{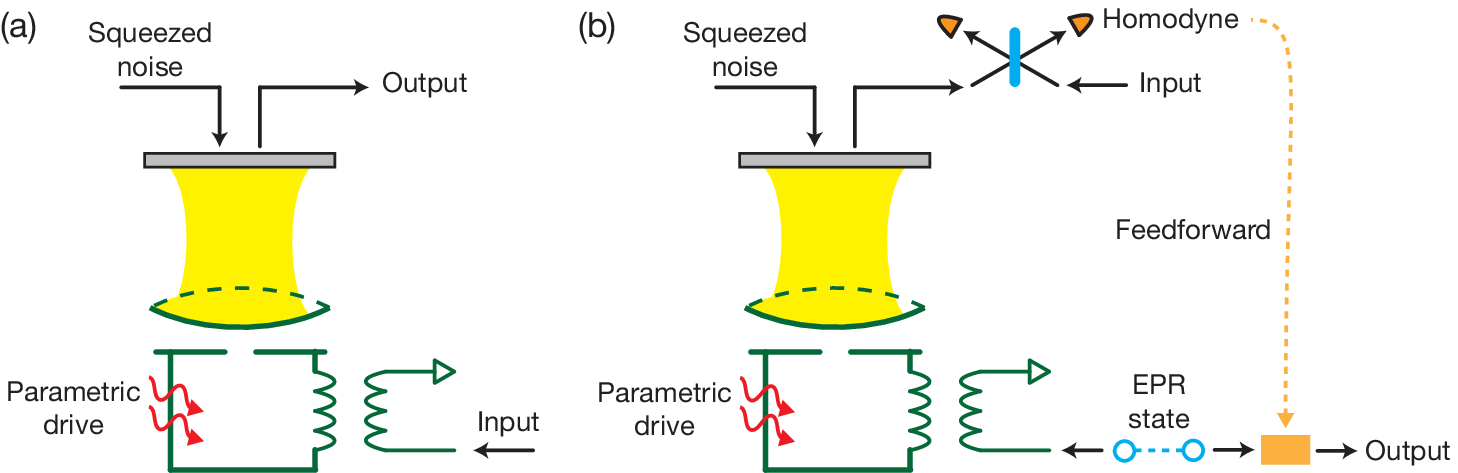}
\caption{Microwave-optics transduction strategy without optical parametric drive.
(a)  Microwave-to-optics transduction.  The microwave input cavity is parametrically driven, and squeezed vacuum is injected to the optical output cavity.  
(b)  Optics-to-microwave transduction.  Instead of feeding the optical signal into the transducer, a microwave EPR state (or, in practice, two-mode-squeezed state) is created in the microwave circuit.  Half of the EPR state is fed into the microwave-to-optics transducer.  The optical signal and the output from the transducer pass through a 50:50 beam splitter and homodyne detected; this implements the continuous-variable Bell measurement.  After feedforward (linear displacement) of the other half of the microwave EPR state, the optical input signal is teleported as a microwave signal.
\label{fig:teleport}}
\end{center}
\end{figure}

Our transduction strategy requires parametrically driving the input cavity and injecting squeezing into the output cavity.  For optics-to-microwave transduction, this requires parametrically driving the optical cavity, which might be challenging in some platforms.  Nevertheless, optical parametric drive is not necessary if we combine microwave-to-optics transducer (see Fig.~\ref{fig:teleport}a) and continuous-variable quantum teleportation.

The teleportation-based strategy is outlined in Fig.~\ref{fig:teleport}b.  An itinerant two-mode-squeezed state, which is an approximated maximally entangled EPR state \cite{Braunstein:2005wr}, is prepared in the microwave transmission line, i.e.
\begin{equation}
\Atel_1[\w] = \cosh r_t \Ain_1[\w] + \sinh r_t \Aind_0[\w]~~,~~\Atel_0[\w] = \sinh r_t \Aind_1[\w] + \cosh r_t \Ain_0[\w]~.
\end{equation}
One half of the two-mode-squeezed state (port 1) is fed into the microwave-to-optics transducer, while the other half (port 0) will be the target of teleportation.  The microwave-to-optics transducer is again described by Eq.~(\ref{eq:trans13}), i.e.
\begin{equation}
\Aout_2[\w]= T_{2,1}(\w)\Atel_1[\w] + T_{2,1^\dag}(\w) \Ateld_1[\w] + T_{2,2}(\w) \Ain_2[\w]  + T_{2,2^\dag}(\w) \Aind_2[\w] ~.
\end{equation}
The optical output from the transducer (port 2) will pass through a 50:50 beam splitter with the optical input signal (port 3).  We define the output of the beam splitter as port 4 and 5, i.e.
\begin{equation}
\Aout_4 =\frac{1}{\sqrt{2}}( e^{i\phi_t}\Aout_2 + \Ain_3 )~~,~~\Aout_5 =\frac{1}{\sqrt{2}}( e^{i\phi_t}\Aout_2 - \Ain_3 )~.
\end{equation}
Both ports are then homodyne detected.  According to detection results, linear displacement is applied on the port 0 (feedforward).  Without loss of generality, we assume port 4 is measured in $Q$-quadrature (i.e. $\propto \Aout_4+\Aoutd_4$) and port 5 is measured in $P$-quadrature (i.e. $\propto \Aout_5 - \Aoutd_5$).

Our goal is to teleport the optical signal contained in the frequency mode $\Ain_3[-\w]$ to the microwave frequency mode $\Atel_0[-\w]$.  By choosing a feedforward strength that yields unit conversion efficiency, the resultant microwave output at port 0 is given by
\begin{eqnarray}
\Atel_0[-\w] &=& \Ain_3[-\w] +(\cosh r_t + e^{i \phi_t} T^\ast_{2,1}(\w) \sinh r_t) \Ain_0[-\w] 
+ (\sinh r_t + e^{i \phi_t} T^\ast_{2,1}(\w) \cosh r_t) \Aind_1[-\w] \nonumber \\
&&+ e^{i\phi_t} T^\ast_{2,1^\dag}(\w) (\cosh r_t \Ain_1[-\w] + \sinh r_t \Aind_0[-\w])
+ e^{i\phi_t} (T^\ast_{2,2^\dag}(\w) \Ain_2[-\w] + T^\ast_{2,2}(\w) \Aind_2[-\w] )~.
\end{eqnarray}
Following Eq.~(\ref{eq:S_def}), we can compute the added noise as
\begin{eqnarray}
S(\w) &=& |\cosh r_t + e^{i \phi_t} T^\ast_{2,1}(\w) \sinh r_t|^2 + |\sinh r_t + e^{i \phi_t} T^\ast_{2,1}(\w) \cosh r_t|^2 + |T^\ast_{2,1^\dag}(\w)|^2 (\cosh^2 r_t + \sinh^2 r_t)  \nonumber \\
&&+ |e^{i\vartheta }\sinh s T_{2,2}(\w) + \cosh s T_{2,2^\dag} (\w)|^2 + \frac{1}{2} (|T_{2,2}(\w)|^2 - |T_{2,2^\dag}(\w)|^2)~.
\end{eqnarray}

We show the typical performance of the teleportation-based strategy in Fig.~\ref{fig:Transduction}b (red dotted line).  The squeezing strength of both the optical cavity injection and the microwave EPR state are picked as 10dB.  The phase of the injected optical squeezing and beam splitter are picked to minimize the added noise at $\w=\w_0$, i.e.
\begin{equation}
e^{i\vartheta}  = -T_{2,2^\dag}(\w_0)/T_{2,2}(\w_0)~~,~~ e^{i \phi_t} = -T_{2,1}(\w_0)/|T_{2,1}(\w_0)|~.
\end{equation}
We note that in this case the added noise is mainly contributed by the finite squeezing of the microwave EPR state.  We also note that adaptive transduction strategies other than teleportation are also available \cite{2009PhRvA..80b2304F, 2018PhRvL.120b0502Z, Transduction_interference}.

\bibliographystyle{apsrev4-1}
\pagestyle{plain}
\bibliography{evade_ref}

\begin{thebibliography}{58}%
\makeatletter
\providecommand \@ifxundefined [1]{%
 \@ifx{#1\undefined}
}%
\providecommand \@ifnum [1]{%
 \ifnum #1\expandafter \@firstoftwo
 \else \expandafter \@secondoftwo
 \fi
}%
\providecommand \@ifx [1]{%
 \ifx #1\expandafter \@firstoftwo
 \else \expandafter \@secondoftwo
 \fi
}%
\providecommand \natexlab [1]{#1}%
\providecommand \enquote  [1]{``#1''}%
\providecommand \bibnamefont  [1]{#1}%
\providecommand \bibfnamefont [1]{#1}%
\providecommand \citenamefont [1]{#1}%
\providecommand \href@noop [0]{\@secondoftwo}%
\providecommand \href [0]{\begingroup \@sanitize@url \@href}%
\providecommand \@href[1]{\@@startlink{#1}\@@href}%
\providecommand \@@href[1]{\endgroup#1\@@endlink}%
\providecommand \@sanitize@url [0]{\catcode `\\12\catcode `\$12\catcode
  `\&12\catcode `\#12\catcode `\^12\catcode `\_12\catcode `\%12\relax}%
\providecommand \@@startlink[1]{}%
\providecommand \@@endlink[0]{}%
\providecommand \url  [0]{\begingroup\@sanitize@url \@url }%
\providecommand \@url [1]{\endgroup\@href {#1}{\urlprefix }}%
\providecommand \urlprefix  [0]{URL }%
\providecommand \Eprint [0]{\href }%
\providecommand \doibase [0]{http://dx.doi.org/}%
\providecommand \selectlanguage [0]{\@gobble}%
\providecommand \bibinfo  [0]{\@secondoftwo}%
\providecommand \bibfield  [0]{\@secondoftwo}%
\providecommand \translation [1]{[#1]}%
\providecommand \BibitemOpen [0]{}%
\providecommand \bibitemStop [0]{}%
\providecommand \bibitemNoStop [0]{.\EOS\space}%
\providecommand \EOS [0]{\spacefactor3000\relax}%
\providecommand \BibitemShut  [1]{\csname bibitem#1\endcsname}%
\let\auto@bib@innerbib\@empty
\bibitem [{\citenamefont {Law}(1995)}]{1995PhRvA..51.2537L}%
  \BibitemOpen
  \bibfield  {author} {\bibinfo {author} {\bibfnamefont {C.~K.}\ \bibnamefont
  {Law}},\ }\href@noop {} {\bibfield  {journal} {\bibinfo  {journal} {Physical
  Review A}\ }\textbf {\bibinfo {volume} {51}},\ \bibinfo {pages} {2537}
  (\bibinfo {year} {1995})}\BibitemShut {NoStop}%
\bibitem [{\citenamefont {Aspelmeyer}\ \emph {et~al.}(2014)\citenamefont
  {Aspelmeyer}, \citenamefont {Kippenberg},\ and\ \citenamefont
  {Marquardt}}]{2014RvMP...86.1391A}%
  \BibitemOpen
  \bibfield  {author} {\bibinfo {author} {\bibfnamefont {M.}~\bibnamefont
  {Aspelmeyer}}, \bibinfo {author} {\bibfnamefont {T.~J.}\ \bibnamefont
  {Kippenberg}}, \ and\ \bibinfo {author} {\bibfnamefont {F.}~\bibnamefont
  {Marquardt}},\ }\href@noop {} {\bibfield  {journal} {\bibinfo  {journal}
  {Rev. Mod. Phys.}\ }\textbf {\bibinfo {volume} {86}},\ \bibinfo {pages}
  {1391} (\bibinfo {year} {2014})}\BibitemShut {NoStop}%
\bibitem [{\citenamefont {Teufel}\ \emph {et~al.}(2011)\citenamefont {Teufel},
  \citenamefont {Donner}, \citenamefont {Li}, \citenamefont {Harlow},
  \citenamefont {Allman}, \citenamefont {Cicak}, \citenamefont {Sirois},
  \citenamefont {Whittaker}, \citenamefont {Lehnert},\ and\ \citenamefont
  {Simmonds}}]{Teufel:2011jg}%
  \BibitemOpen
  \bibfield  {author} {\bibinfo {author} {\bibfnamefont {J.~D.}\ \bibnamefont
  {Teufel}}, \bibinfo {author} {\bibfnamefont {T.}~\bibnamefont {Donner}},
  \bibinfo {author} {\bibfnamefont {D.}~\bibnamefont {Li}}, \bibinfo {author}
  {\bibfnamefont {J.~W.}\ \bibnamefont {Harlow}}, \bibinfo {author}
  {\bibfnamefont {M.~S.}\ \bibnamefont {Allman}}, \bibinfo {author}
  {\bibfnamefont {K.}~\bibnamefont {Cicak}}, \bibinfo {author} {\bibfnamefont
  {A.~J.}\ \bibnamefont {Sirois}}, \bibinfo {author} {\bibfnamefont {J.~D.}\
  \bibnamefont {Whittaker}}, \bibinfo {author} {\bibfnamefont {K.~W.}\
  \bibnamefont {Lehnert}}, \ and\ \bibinfo {author} {\bibfnamefont {R.~W.}\
  \bibnamefont {Simmonds}},\ }\href@noop {} {\bibfield  {journal} {\bibinfo
  {journal} {Nature}\ }\textbf {\bibinfo {volume} {475}},\ \bibinfo {pages}
  {359} (\bibinfo {year} {2011})}\BibitemShut {NoStop}%
\bibitem [{\citenamefont {Chan}\ \emph {et~al.}(2011)\citenamefont {Chan},
  \citenamefont {Alegre}, \citenamefont {Safavi-Naeini}, \citenamefont {Hill},
  \citenamefont {Krause}, \citenamefont {Gr{\"o}blacher}, \citenamefont
  {Aspelmeyer},\ and\ \citenamefont {Painter}}]{Chan:2011dy}%
  \BibitemOpen
  \bibfield  {author} {\bibinfo {author} {\bibfnamefont {J.}~\bibnamefont
  {Chan}}, \bibinfo {author} {\bibfnamefont {T.~P.~M.}\ \bibnamefont {Alegre}},
  \bibinfo {author} {\bibfnamefont {A.~H.}\ \bibnamefont {Safavi-Naeini}},
  \bibinfo {author} {\bibfnamefont {J.~T.}\ \bibnamefont {Hill}}, \bibinfo
  {author} {\bibfnamefont {A.}~\bibnamefont {Krause}}, \bibinfo {author}
  {\bibfnamefont {S.}~\bibnamefont {Gr{\"o}blacher}}, \bibinfo {author}
  {\bibfnamefont {M.}~\bibnamefont {Aspelmeyer}}, \ and\ \bibinfo {author}
  {\bibfnamefont {O.}~\bibnamefont {Painter}},\ }\href@noop {} {\bibfield
  {journal} {\bibinfo  {journal} {Nature}\ }\textbf {\bibinfo {volume} {478}},\
  \bibinfo {pages} {89} (\bibinfo {year} {2011})}\BibitemShut {NoStop}%
\bibitem [{\citenamefont {Safavi-Naeini}\ and\ \citenamefont
  {Painter}(2011)}]{2011NJPh...13a3017S}%
  \BibitemOpen
  \bibfield  {author} {\bibinfo {author} {\bibfnamefont {A.~H.}\ \bibnamefont
  {Safavi-Naeini}}\ and\ \bibinfo {author} {\bibfnamefont {O.}~\bibnamefont
  {Painter}},\ }\href@noop {} {\bibfield  {journal} {\bibinfo  {journal} {New
  Journal of Physics}\ }\textbf {\bibinfo {volume} {13}},\ \bibinfo {pages}
  {013017} (\bibinfo {year} {2011})}\BibitemShut {NoStop}%
\bibitem [{\citenamefont {Regal}\ and\ \citenamefont
  {Lehnert}(2011)}]{2011JPhCS.264a2025R}%
  \BibitemOpen
  \bibfield  {author} {\bibinfo {author} {\bibfnamefont {C.~A.}\ \bibnamefont
  {Regal}}\ and\ \bibinfo {author} {\bibfnamefont {K.~W.}\ \bibnamefont
  {Lehnert}},\ }\href@noop {} {\bibfield  {journal} {\bibinfo  {journal}
  {Journal of Physics: Conference Series}\ }\textbf {\bibinfo {volume} {264}},\
  \bibinfo {pages} {012025} (\bibinfo {year} {2011})}\BibitemShut {NoStop}%
\bibitem [{\citenamefont {Andrews}\ \emph {et~al.}(2014)\citenamefont
  {Andrews}, \citenamefont {Peterson}, \citenamefont {Purdy}, \citenamefont
  {Cicak}, \citenamefont {Simmonds}, \citenamefont {Regal},\ and\ \citenamefont
  {Lehnert}}]{2014NatPh..10..321A}%
  \BibitemOpen
  \bibfield  {author} {\bibinfo {author} {\bibfnamefont {R.~W.}\ \bibnamefont
  {Andrews}}, \bibinfo {author} {\bibfnamefont {R.~W.}\ \bibnamefont
  {Peterson}}, \bibinfo {author} {\bibfnamefont {T.~P.}\ \bibnamefont {Purdy}},
  \bibinfo {author} {\bibfnamefont {K.}~\bibnamefont {Cicak}}, \bibinfo
  {author} {\bibfnamefont {R.~W.}\ \bibnamefont {Simmonds}}, \bibinfo {author}
  {\bibfnamefont {C.~A.}\ \bibnamefont {Regal}}, \ and\ \bibinfo {author}
  {\bibfnamefont {K.~W.}\ \bibnamefont {Lehnert}},\ }\href@noop {} {\bibfield
  {journal} {\bibinfo  {journal} {Nature Physics}\ }\textbf {\bibinfo {volume}
  {10}},\ \bibinfo {pages} {321} (\bibinfo {year} {2014})}\BibitemShut
  {NoStop}%
\bibitem [{\citenamefont {Higginbotham}\ \emph {et~al.}(2018)\citenamefont
  {Higginbotham}, \citenamefont {Burns}, \citenamefont {Urmey}, \citenamefont
  {Peterson}, \citenamefont {Kampel}, \citenamefont {Brubaker}, \citenamefont
  {Smith}, \citenamefont {Lehnert},\ and\ \citenamefont
  {Regal}}]{Higginbotham:2018ca}%
  \BibitemOpen
  \bibfield  {author} {\bibinfo {author} {\bibfnamefont {A.~P.}\ \bibnamefont
  {Higginbotham}}, \bibinfo {author} {\bibfnamefont {P.~S.}\ \bibnamefont
  {Burns}}, \bibinfo {author} {\bibfnamefont {M.~D.}\ \bibnamefont {Urmey}},
  \bibinfo {author} {\bibfnamefont {R.~W.}\ \bibnamefont {Peterson}}, \bibinfo
  {author} {\bibfnamefont {N.~S.}\ \bibnamefont {Kampel}}, \bibinfo {author}
  {\bibfnamefont {B.~M.}\ \bibnamefont {Brubaker}}, \bibinfo {author}
  {\bibfnamefont {G.}~\bibnamefont {Smith}}, \bibinfo {author} {\bibfnamefont
  {K.~W.}\ \bibnamefont {Lehnert}}, \ and\ \bibinfo {author} {\bibfnamefont
  {C.~A.}\ \bibnamefont {Regal}},\ }\href@noop {} {\bibfield  {journal}
  {\bibinfo  {journal} {Nature Physics}\ }\textbf {\bibinfo {volume} {14}},\
  \bibinfo {pages} {1038} (\bibinfo {year} {2018})}\BibitemShut {NoStop}%
\bibitem [{\citenamefont {Tian}(2015)}]{2015AnP...527....1T}%
  \BibitemOpen
  \bibfield  {author} {\bibinfo {author} {\bibfnamefont {L.}~\bibnamefont
  {Tian}},\ }\href@noop {} {\bibfield  {journal} {\bibinfo  {journal} {Annalen
  der Physik}\ }\textbf {\bibinfo {volume} {527}},\ \bibinfo {pages} {1}
  (\bibinfo {year} {2015})}\BibitemShut {NoStop}%
\bibitem [{\citenamefont {Kurizki}\ \emph {et~al.}(2015)\citenamefont
  {Kurizki}, \citenamefont {Bertet}, \citenamefont {Kubo}, \citenamefont
  {Molmer}, \citenamefont {Petrosyan}, \citenamefont {Rabl},\ and\
  \citenamefont {Schmiedmayer}}]{2015PNAS..112.3866K}%
  \BibitemOpen
  \bibfield  {author} {\bibinfo {author} {\bibfnamefont {G.}~\bibnamefont
  {Kurizki}}, \bibinfo {author} {\bibfnamefont {P.}~\bibnamefont {Bertet}},
  \bibinfo {author} {\bibfnamefont {Y.}~\bibnamefont {Kubo}}, \bibinfo {author}
  {\bibfnamefont {K.}~\bibnamefont {Molmer}}, \bibinfo {author} {\bibfnamefont
  {D.}~\bibnamefont {Petrosyan}}, \bibinfo {author} {\bibfnamefont
  {P.}~\bibnamefont {Rabl}}, \ and\ \bibinfo {author} {\bibfnamefont
  {J.}~\bibnamefont {Schmiedmayer}},\ }\href@noop {} {\bibfield  {journal}
  {\bibinfo  {journal} {PNAS}\ }\textbf {\bibinfo {volume} {112}},\ \bibinfo
  {pages} {3866} (\bibinfo {year} {2015})}\BibitemShut {NoStop}%
\bibitem [{\citenamefont {Marquardt}\ \emph {et~al.}(2007)\citenamefont
  {Marquardt}, \citenamefont {Chen}, \citenamefont {Clerk},\ and\ \citenamefont
  {Girvin}}]{Marquardt:2007dn}%
  \BibitemOpen
  \bibfield  {author} {\bibinfo {author} {\bibfnamefont {F.}~\bibnamefont
  {Marquardt}}, \bibinfo {author} {\bibfnamefont {J.~P.}\ \bibnamefont {Chen}},
  \bibinfo {author} {\bibfnamefont {A.~A.}\ \bibnamefont {Clerk}}, \ and\
  \bibinfo {author} {\bibfnamefont {S.~M.}\ \bibnamefont {Girvin}},\
  }\href@noop {} {\bibfield  {journal} {\bibinfo  {journal} {Physical Review
  Letters}\ }\textbf {\bibinfo {volume} {99}},\ \bibinfo {pages} {093902}
  (\bibinfo {year} {2007})}\BibitemShut {NoStop}%
\bibitem [{\citenamefont {Wilson-Rae}\ \emph {et~al.}(2007)\citenamefont
  {Wilson-Rae}, \citenamefont {Nooshi}, \citenamefont {Zwerger},\ and\
  \citenamefont {Kippenberg}}]{WilsonRae:2007jp}%
  \BibitemOpen
  \bibfield  {author} {\bibinfo {author} {\bibfnamefont {I.}~\bibnamefont
  {Wilson-Rae}}, \bibinfo {author} {\bibfnamefont {N.}~\bibnamefont {Nooshi}},
  \bibinfo {author} {\bibfnamefont {W.}~\bibnamefont {Zwerger}}, \ and\
  \bibinfo {author} {\bibfnamefont {T.~J.}\ \bibnamefont {Kippenberg}},\
  }\href@noop {} {\bibfield  {journal} {\bibinfo  {journal} {Physical Review
  Letters}\ }\textbf {\bibinfo {volume} {99}},\ \bibinfo {pages} {093901}
  (\bibinfo {year} {2007})}\BibitemShut {NoStop}%
\bibitem [{\citenamefont {Elste}\ \emph {et~al.}(2009)\citenamefont {Elste},
  \citenamefont {Girvin},\ and\ \citenamefont {Clerk}}]{2009PhRvL.102t7209E}%
  \BibitemOpen
  \bibfield  {author} {\bibinfo {author} {\bibfnamefont {F.}~\bibnamefont
  {Elste}}, \bibinfo {author} {\bibfnamefont {S.~M.}\ \bibnamefont {Girvin}}, \
  and\ \bibinfo {author} {\bibfnamefont {A.~A.}\ \bibnamefont {Clerk}},\
  }\href@noop {} {\bibfield  {journal} {\bibinfo  {journal} {Physical Review
  Letters}\ }\textbf {\bibinfo {volume} {102}},\ \bibinfo {pages} {207209}
  (\bibinfo {year} {2009})}\BibitemShut {NoStop}%
\bibitem [{\citenamefont {Xuereb}\ \emph {et~al.}(2011)\citenamefont {Xuereb},
  \citenamefont {Schnabel},\ and\ \citenamefont
  {Hammerer}}]{2011PhRvL.107u3604X}%
  \BibitemOpen
  \bibfield  {author} {\bibinfo {author} {\bibfnamefont {A.}~\bibnamefont
  {Xuereb}}, \bibinfo {author} {\bibfnamefont {R.}~\bibnamefont {Schnabel}}, \
  and\ \bibinfo {author} {\bibfnamefont {K.}~\bibnamefont {Hammerer}},\
  }\href@noop {} {\bibfield  {journal} {\bibinfo  {journal} {Physical Review
  Letters}\ }\textbf {\bibinfo {volume} {107}},\ \bibinfo {pages} {213604}
  (\bibinfo {year} {2011})}\BibitemShut {NoStop}%
\bibitem [{\citenamefont {Sawadsky}\ \emph {et~al.}(2015)\citenamefont
  {Sawadsky}, \citenamefont {Kaufer}, \citenamefont {Nia}, \citenamefont
  {Tarabrin}, \citenamefont {Khalili}, \citenamefont {Hammerer},\ and\
  \citenamefont {Schnabel}}]{2015PhRvL.114d3601S}%
  \BibitemOpen
  \bibfield  {author} {\bibinfo {author} {\bibfnamefont {A.}~\bibnamefont
  {Sawadsky}}, \bibinfo {author} {\bibfnamefont {H.}~\bibnamefont {Kaufer}},
  \bibinfo {author} {\bibfnamefont {R.~M.}\ \bibnamefont {Nia}}, \bibinfo
  {author} {\bibfnamefont {S.~P.}\ \bibnamefont {Tarabrin}}, \bibinfo {author}
  {\bibfnamefont {F.~Y.}\ \bibnamefont {Khalili}}, \bibinfo {author}
  {\bibfnamefont {K.}~\bibnamefont {Hammerer}}, \ and\ \bibinfo {author}
  {\bibfnamefont {R.}~\bibnamefont {Schnabel}},\ }\href@noop {} {\bibfield
  {journal} {\bibinfo  {journal} {Physical Review Letters}\ }\textbf {\bibinfo
  {volume} {114}},\ \bibinfo {pages} {043601} (\bibinfo {year}
  {2015})}\BibitemShut {NoStop}%
\bibitem [{\citenamefont {Huang}\ and\ \citenamefont
  {Chen}(2018)}]{Huang:2018br}%
  \BibitemOpen
  \bibfield  {author} {\bibinfo {author} {\bibfnamefont {S.}~\bibnamefont
  {Huang}}\ and\ \bibinfo {author} {\bibfnamefont {A.}~\bibnamefont {Chen}},\
  }\href@noop {} {\bibfield  {journal} {\bibinfo  {journal} {Physical Review
  A}\ }\textbf {\bibinfo {volume} {98}},\ \bibinfo {pages} {063818} (\bibinfo
  {year} {2018})}\BibitemShut {NoStop}%
\bibitem [{\citenamefont {Genes}\ \emph {et~al.}(2009)\citenamefont {Genes},
  \citenamefont {Ritsch},\ and\ \citenamefont {Vitali}}]{2009PhRvA..80f1803G}%
  \BibitemOpen
  \bibfield  {author} {\bibinfo {author} {\bibfnamefont {C.}~\bibnamefont
  {Genes}}, \bibinfo {author} {\bibfnamefont {H.}~\bibnamefont {Ritsch}}, \
  and\ \bibinfo {author} {\bibfnamefont {D.}~\bibnamefont {Vitali}},\
  }\href@noop {} {\bibfield  {journal} {\bibinfo  {journal} {Physical Review
  A}\ }\textbf {\bibinfo {volume} {80}},\ \bibinfo {pages} {061803(R)}
  (\bibinfo {year} {2009})}\BibitemShut {NoStop}%
\bibitem [{\citenamefont {Genes}\ \emph {et~al.}(2011)\citenamefont {Genes},
  \citenamefont {Ritsch}, \citenamefont {Drewsen},\ and\ \citenamefont
  {Dantan}}]{2011PhRvA..84e1801G}%
  \BibitemOpen
  \bibfield  {author} {\bibinfo {author} {\bibfnamefont {C.}~\bibnamefont
  {Genes}}, \bibinfo {author} {\bibfnamefont {H.}~\bibnamefont {Ritsch}},
  \bibinfo {author} {\bibfnamefont {M.}~\bibnamefont {Drewsen}}, \ and\
  \bibinfo {author} {\bibfnamefont {A.}~\bibnamefont {Dantan}},\ }\href@noop {}
  {\bibfield  {journal} {\bibinfo  {journal} {Physical Review A}\ }\textbf
  {\bibinfo {volume} {84}},\ \bibinfo {pages} {051801(R)} (\bibinfo {year}
  {2011})}\BibitemShut {NoStop}%
\bibitem [{\citenamefont {Lau}\ \emph {et~al.}(2018)\citenamefont {Lau},
  \citenamefont {Eisfeld},\ and\ \citenamefont {Rost}}]{Lau:2018ev}%
  \BibitemOpen
  \bibfield  {author} {\bibinfo {author} {\bibfnamefont {H.-K.}\ \bibnamefont
  {Lau}}, \bibinfo {author} {\bibfnamefont {A.}~\bibnamefont {Eisfeld}}, \ and\
  \bibinfo {author} {\bibfnamefont {J.-M.}\ \bibnamefont {Rost}},\ }\href@noop
  {} {\bibfield  {journal} {\bibinfo  {journal} {Physical Review A}\ }\textbf
  {\bibinfo {volume} {98}},\ \bibinfo {pages} {043827} (\bibinfo {year}
  {2018})}\BibitemShut {NoStop}%
\bibitem [{\citenamefont {Huang}\ and\ \citenamefont
  {Agarwal}(2009)}]{2009PhRvA..79a3821H}%
  \BibitemOpen
  \bibfield  {author} {\bibinfo {author} {\bibfnamefont {S.}~\bibnamefont
  {Huang}}\ and\ \bibinfo {author} {\bibfnamefont {G.~S.}\ \bibnamefont
  {Agarwal}},\ }\href@noop {} {\bibfield  {journal} {\bibinfo  {journal}
  {Physical Review A}\ }\textbf {\bibinfo {volume} {79}},\ \bibinfo {pages}
  {013821} (\bibinfo {year} {2009})}\BibitemShut {NoStop}%
\bibitem [{\citenamefont {Asghari~Nejad}\ \emph {et~al.}(2018)\citenamefont
  {Asghari~Nejad}, \citenamefont {Askari},\ and\ \citenamefont
  {Baghshahi}}]{2018PhyE..102...83A}%
  \BibitemOpen
  \bibfield  {author} {\bibinfo {author} {\bibfnamefont {A.}~\bibnamefont
  {Asghari~Nejad}}, \bibinfo {author} {\bibfnamefont {H.~R.}\ \bibnamefont
  {Askari}}, \ and\ \bibinfo {author} {\bibfnamefont {H.~R.}\ \bibnamefont
  {Baghshahi}},\ }\href@noop {} {\bibfield  {journal} {\bibinfo  {journal}
  {Physica E: Low-dimensional Systems and Nanostructures}\ }\textbf {\bibinfo
  {volume} {102}},\ \bibinfo {pages} {83} (\bibinfo {year} {2018})}\BibitemShut
  {NoStop}%
\bibitem [{\citenamefont {Clark}\ \emph {et~al.}(2017)\citenamefont {Clark},
  \citenamefont {Lecocq}, \citenamefont {Simmonds}, \citenamefont {Aumentado},\
  and\ \citenamefont {Teufel}}]{2017Natur.541..191C}%
  \BibitemOpen
  \bibfield  {author} {\bibinfo {author} {\bibfnamefont {J.~B.}\ \bibnamefont
  {Clark}}, \bibinfo {author} {\bibfnamefont {F.}~\bibnamefont {Lecocq}},
  \bibinfo {author} {\bibfnamefont {R.~W.}\ \bibnamefont {Simmonds}}, \bibinfo
  {author} {\bibfnamefont {J.}~\bibnamefont {Aumentado}}, \ and\ \bibinfo
  {author} {\bibfnamefont {J.~D.}\ \bibnamefont {Teufel}},\ }\href@noop {}
  {\bibfield  {journal} {\bibinfo  {journal} {Nature}\ }\textbf {\bibinfo
  {volume} {541}},\ \bibinfo {pages} {191} (\bibinfo {year}
  {2017})}\BibitemShut {NoStop}%
\bibitem [{\citenamefont {Asjad}\ \emph {et~al.}(2016)\citenamefont {Asjad},
  \citenamefont {Zippilli},\ and\ \citenamefont
  {Vitali}}]{2016PhRvA..94e1801A}%
  \BibitemOpen
  \bibfield  {author} {\bibinfo {author} {\bibfnamefont {M.}~\bibnamefont
  {Asjad}}, \bibinfo {author} {\bibfnamefont {S.}~\bibnamefont {Zippilli}}, \
  and\ \bibinfo {author} {\bibfnamefont {D.}~\bibnamefont {Vitali}},\
  }\href@noop {} {\bibfield  {journal} {\bibinfo  {journal} {Physical Review
  A}\ }\textbf {\bibinfo {volume} {94}},\ \bibinfo {pages} {051801(R)}
  (\bibinfo {year} {2016})}\BibitemShut {NoStop}%
\bibitem [{Note1()}]{Note1}%
  \BibitemOpen
  \bibinfo {note} {Note that the mechanical mode and bath are unaffected by the
  diagonalization, i.e. $ \protect \mathaccentV {hat}05E{\alpha } _\protect
  \textrm {m} = \protect \mathaccentV {hat}05E{a} _\protect \textrm {m}$ and $
  \protect \mathaccentV {hat}05E{\protect \mathcal {A}}^\protect \textrm {in}
  _\protect \textrm {m} = \protect \mathaccentV {hat}05E{A}^\protect \textrm
  {in} _\protect \textrm {m}$.}\BibitemShut {Stop}%
\bibitem [{\citenamefont {Weedbrook}\ \emph {et~al.}(2012)\citenamefont
  {Weedbrook}, \citenamefont {Pirandola}, \citenamefont {Garc\'{\i}a-Patr\'on},
  \citenamefont {Cerf}, \citenamefont {Ralph}, \citenamefont {Shapiro},\ and\
  \citenamefont {Lloyd}}]{Weedbrook:2012fe}%
  \BibitemOpen
  \bibfield  {author} {\bibinfo {author} {\bibfnamefont {C.}~\bibnamefont
  {Weedbrook}}, \bibinfo {author} {\bibfnamefont {S.}~\bibnamefont
  {Pirandola}}, \bibinfo {author} {\bibfnamefont {R.}~\bibnamefont
  {Garc\'{\i}a-Patr\'on}}, \bibinfo {author} {\bibfnamefont {N.~J.}\
  \bibnamefont {Cerf}}, \bibinfo {author} {\bibfnamefont {T.~C.}\ \bibnamefont
  {Ralph}}, \bibinfo {author} {\bibfnamefont {J.~H.}\ \bibnamefont {Shapiro}},
  \ and\ \bibinfo {author} {\bibfnamefont {S.}~\bibnamefont {Lloyd}},\
  }\href@noop {} {\bibfield  {journal} {\bibinfo  {journal} {Rev. Mod. Phys.}\
  }\textbf {\bibinfo {volume} {84}},\ \bibinfo {pages} {621} (\bibinfo {year}
  {2012})}\BibitemShut {NoStop}%
\bibitem [{see()}]{seeSI}%
  \BibitemOpen
  \bibinfo {note} {More details can be found in Supplementary Material, which
  also contains Ref.~\cite{2015PhRvA..91a3807W, 1987PhRvA..35.5288D,
  Peterson:2019, 2009PhRvA..80b2304F, 2018PhRvL.120b0502Z}}\BibitemShut
  {NoStop}%
\bibitem [{\citenamefont {Milburn}\ and\ \citenamefont
  {Walls}(1981)}]{MILBURN:1981tu}%
  \BibitemOpen
  \bibfield  {author} {\bibinfo {author} {\bibfnamefont {G.}~\bibnamefont
  {Milburn}}\ and\ \bibinfo {author} {\bibfnamefont {D.~F.}\ \bibnamefont
  {Walls}},\ }\href@noop {} {\bibfield  {journal} {\bibinfo  {journal} {Optics
  Communications}\ }\textbf {\bibinfo {volume} {39}},\ \bibinfo {pages} {401}
  (\bibinfo {year} {1981})}\BibitemShut {NoStop}%
\bibitem [{\citenamefont {Caves}(1982)}]{1982PhRvD..26.1817C}%
  \BibitemOpen
  \bibfield  {author} {\bibinfo {author} {\bibfnamefont {C.~M.}\ \bibnamefont
  {Caves}},\ }\href@noop {} {\bibfield  {journal} {\bibinfo  {journal}
  {Physical Review D}\ }\textbf {\bibinfo {volume} {26}},\ \bibinfo {pages}
  {1817} (\bibinfo {year} {1982})}\BibitemShut {NoStop}%
\bibitem [{\citenamefont {Braunstein}\ and\ \citenamefont {van
  Loock}(2005)}]{Braunstein:2005wr}%
  \BibitemOpen
  \bibfield  {author} {\bibinfo {author} {\bibfnamefont {S.~L.}\ \bibnamefont
  {Braunstein}}\ and\ \bibinfo {author} {\bibfnamefont {P.}~\bibnamefont {van
  Loock}},\ }\href@noop {} {\bibfield  {journal} {\bibinfo  {journal} {Reviews
  of Modern Physics}\ }\textbf {\bibinfo {volume} {77}},\ \bibinfo {pages}
  {513} (\bibinfo {year} {2005})}\BibitemShut {NoStop}%
\bibitem [{\citenamefont {Lau}\ and\ \citenamefont
  {Clerk}(2019)}]{Transduction_interference}%
  \BibitemOpen
  \bibfield  {author} {\bibinfo {author} {\bibfnamefont {H.-K.}\ \bibnamefont
  {Lau}}\ and\ \bibinfo {author} {\bibfnamefont {A.~A.}\ \bibnamefont
  {Clerk}},\ }\href@noop {} {\bibfield  {journal} {\bibinfo  {journal} {npj
  Quantum Information}\ }\textbf {\bibinfo {volume} {5}},\ \bibinfo {pages}
  {31} (\bibinfo {year} {2019})}\BibitemShut {NoStop}%
\bibitem [{\citenamefont {Yamamoto}\ \emph {et~al.}(2008)\citenamefont
  {Yamamoto}, \citenamefont {Inomata}, \citenamefont {Watanabe}, \citenamefont
  {Matsuba}, \citenamefont {Miyazaki}, \citenamefont {Oliver}, \citenamefont
  {Nakamura},\ and\ \citenamefont {Tsai}}]{2008ApPhL..93d2510Y}%
  \BibitemOpen
  \bibfield  {author} {\bibinfo {author} {\bibfnamefont {T.}~\bibnamefont
  {Yamamoto}}, \bibinfo {author} {\bibfnamefont {K.}~\bibnamefont {Inomata}},
  \bibinfo {author} {\bibfnamefont {M.}~\bibnamefont {Watanabe}}, \bibinfo
  {author} {\bibfnamefont {K.}~\bibnamefont {Matsuba}}, \bibinfo {author}
  {\bibfnamefont {T.}~\bibnamefont {Miyazaki}}, \bibinfo {author}
  {\bibfnamefont {W.~D.}\ \bibnamefont {Oliver}}, \bibinfo {author}
  {\bibfnamefont {Y.}~\bibnamefont {Nakamura}}, \ and\ \bibinfo {author}
  {\bibfnamefont {J.~S.}\ \bibnamefont {Tsai}},\ }\href@noop {} {\bibfield
  {journal} {\bibinfo  {journal} {Applied Physics Letters}\ }\textbf {\bibinfo
  {volume} {93}},\ \bibinfo {pages} {042510} (\bibinfo {year}
  {2008})}\BibitemShut {NoStop}%
\bibitem [{\citenamefont {Castellanos-Beltran}\ \emph
  {et~al.}(2008)\citenamefont {Castellanos-Beltran}, \citenamefont {Irwin},
  \citenamefont {Hilton}, \citenamefont {Vale},\ and\ \citenamefont
  {Lehnert}}]{2008NatPh...4..929C}%
  \BibitemOpen
  \bibfield  {author} {\bibinfo {author} {\bibfnamefont {M.~A.}\ \bibnamefont
  {Castellanos-Beltran}}, \bibinfo {author} {\bibfnamefont {K.~D.}\
  \bibnamefont {Irwin}}, \bibinfo {author} {\bibfnamefont {G.~C.}\ \bibnamefont
  {Hilton}}, \bibinfo {author} {\bibfnamefont {L.~R.}\ \bibnamefont {Vale}}, \
  and\ \bibinfo {author} {\bibfnamefont {K.~W.}\ \bibnamefont {Lehnert}},\
  }\href@noop {} {\bibfield  {journal} {\bibinfo  {journal} {Nature Physics}\
  }\textbf {\bibinfo {volume} {4}},\ \bibinfo {pages} {929} (\bibinfo {year}
  {2008})}\BibitemShut {NoStop}%
\bibitem [{\citenamefont {F{\"u}rst}\ \emph {et~al.}(2011)\citenamefont
  {F{\"u}rst}, \citenamefont {Strekalov}, \citenamefont {Elser}, \citenamefont
  {Aiello}, \citenamefont {Andersen}, \citenamefont {Marquardt},\ and\
  \citenamefont {Leuchs}}]{2011PhRvL.106k3901F}%
  \BibitemOpen
  \bibfield  {author} {\bibinfo {author} {\bibfnamefont {J.~U.}\ \bibnamefont
  {F{\"u}rst}}, \bibinfo {author} {\bibfnamefont {D.~V.}\ \bibnamefont
  {Strekalov}}, \bibinfo {author} {\bibfnamefont {D.}~\bibnamefont {Elser}},
  \bibinfo {author} {\bibfnamefont {A.}~\bibnamefont {Aiello}}, \bibinfo
  {author} {\bibfnamefont {U.~L.}\ \bibnamefont {Andersen}}, \bibinfo {author}
  {\bibfnamefont {C.}~\bibnamefont {Marquardt}}, \ and\ \bibinfo {author}
  {\bibfnamefont {G.}~\bibnamefont {Leuchs}},\ }\href@noop {} {\bibfield
  {journal} {\bibinfo  {journal} {Physical Review Letters}\ }\textbf {\bibinfo
  {volume} {106}},\ \bibinfo {pages} {113901} (\bibinfo {year}
  {2011})}\BibitemShut {NoStop}%
\bibitem [{\citenamefont {Vahlbruch}\ \emph {et~al.}(2016)\citenamefont
  {Vahlbruch}, \citenamefont {Mehmet}, \citenamefont {Danzmann},\ and\
  \citenamefont {Schnabel}}]{2016PhRvL.117k0801V}%
  \BibitemOpen
  \bibfield  {author} {\bibinfo {author} {\bibfnamefont {H.}~\bibnamefont
  {Vahlbruch}}, \bibinfo {author} {\bibfnamefont {M.}~\bibnamefont {Mehmet}},
  \bibinfo {author} {\bibfnamefont {K.}~\bibnamefont {Danzmann}}, \ and\
  \bibinfo {author} {\bibfnamefont {R.}~\bibnamefont {Schnabel}},\ }\href@noop
  {} {\bibfield  {journal} {\bibinfo  {journal} {Physical Review Letters}\
  }\textbf {\bibinfo {volume} {117}},\ \bibinfo {pages} {110801} (\bibinfo
  {year} {2016})}\BibitemShut {NoStop}%
\bibitem [{\citenamefont {Peano}\ \emph {et~al.}(2015)\citenamefont {Peano},
  \citenamefont {Schwefel}, \citenamefont {Marquardt},\ and\ \citenamefont
  {Marquardt}}]{2015PhRvL.115x3603P}%
  \BibitemOpen
  \bibfield  {author} {\bibinfo {author} {\bibfnamefont {V.}~\bibnamefont
  {Peano}}, \bibinfo {author} {\bibfnamefont {H.~G.~L.}\ \bibnamefont
  {Schwefel}}, \bibinfo {author} {\bibfnamefont {C.}~\bibnamefont {Marquardt}},
  \ and\ \bibinfo {author} {\bibfnamefont {F.}~\bibnamefont {Marquardt}},\
  }\href@noop {} {\bibfield  {journal} {\bibinfo  {journal} {Physical Review
  Letters}\ }\textbf {\bibinfo {volume} {115}},\ \bibinfo {pages} {243603}
  (\bibinfo {year} {2015})}\BibitemShut {NoStop}%
\bibitem [{\citenamefont {L{\"u}}\ \emph {et~al.}(2015)\citenamefont {L{\"u}},
  \citenamefont {Wu}, \citenamefont {Johansson}, \citenamefont {Jing},
  \citenamefont {Zhang},\ and\ \citenamefont {Nori}}]{2015PhRvL.114i3602L}%
  \BibitemOpen
  \bibfield  {author} {\bibinfo {author} {\bibfnamefont {X.-Y.}\ \bibnamefont
  {L{\"u}}}, \bibinfo {author} {\bibfnamefont {Y.}~\bibnamefont {Wu}}, \bibinfo
  {author} {\bibfnamefont {J.~R.}\ \bibnamefont {Johansson}}, \bibinfo {author}
  {\bibfnamefont {H.}~\bibnamefont {Jing}}, \bibinfo {author} {\bibfnamefont
  {J.}~\bibnamefont {Zhang}}, \ and\ \bibinfo {author} {\bibfnamefont
  {F.}~\bibnamefont {Nori}},\ }\href@noop {} {\bibfield  {journal} {\bibinfo
  {journal} {Physical Review Letters}\ }\textbf {\bibinfo {volume} {114}},\
  \bibinfo {pages} {093602} (\bibinfo {year} {2015})}\BibitemShut {NoStop}%
\bibitem [{\citenamefont {Leroux}\ \emph {et~al.}(2018)\citenamefont {Leroux},
  \citenamefont {Govia},\ and\ \citenamefont {Clerk}}]{Leroux:2018kj}%
  \BibitemOpen
  \bibfield  {author} {\bibinfo {author} {\bibfnamefont {C.}~\bibnamefont
  {Leroux}}, \bibinfo {author} {\bibfnamefont {L.~C.~G.}\ \bibnamefont
  {Govia}}, \ and\ \bibinfo {author} {\bibfnamefont {A.~A.}\ \bibnamefont
  {Clerk}},\ }\href@noop {} {\bibfield  {journal} {\bibinfo  {journal}
  {Physical Review Letters}\ }\textbf {\bibinfo {volume} {120}},\ \bibinfo
  {pages} {093602} (\bibinfo {year} {2018})}\BibitemShut {NoStop}%
\bibitem [{\citenamefont {Qin}\ \emph {et~al.}(2018)\citenamefont {Qin},
  \citenamefont {Miranowicz}, \citenamefont {Li}, \citenamefont {L{\"u}},
  \citenamefont {You},\ and\ \citenamefont {Nori}}]{2018PhRvL.120i3601Q}%
  \BibitemOpen
  \bibfield  {author} {\bibinfo {author} {\bibfnamefont {W.}~\bibnamefont
  {Qin}}, \bibinfo {author} {\bibfnamefont {A.}~\bibnamefont {Miranowicz}},
  \bibinfo {author} {\bibfnamefont {P.-B.}\ \bibnamefont {Li}}, \bibinfo
  {author} {\bibfnamefont {X.-Y.}\ \bibnamefont {L{\"u}}}, \bibinfo {author}
  {\bibfnamefont {J.~Q.}\ \bibnamefont {You}}, \ and\ \bibinfo {author}
  {\bibfnamefont {F.}~\bibnamefont {Nori}},\ }\href@noop {} {\bibfield
  {journal} {\bibinfo  {journal} {Physical Review Letters}\ }\textbf {\bibinfo
  {volume} {120}},\ \bibinfo {pages} {093601} (\bibinfo {year}
  {2018})}\BibitemShut {NoStop}%
\bibitem [{\citenamefont {Vahlbruch}\ \emph {et~al.}(2008)\citenamefont
  {Vahlbruch}, \citenamefont {Mehmet}, \citenamefont {Chelkowski},
  \citenamefont {Hage}, \citenamefont {Franzen}, \citenamefont {Lastzka},
  \citenamefont {Go{\ss}ler}, \citenamefont {Danzmann},\ and\ \citenamefont
  {Schnabel}}]{2008PhRvL.100c3602V}%
  \BibitemOpen
  \bibfield  {author} {\bibinfo {author} {\bibfnamefont {H.}~\bibnamefont
  {Vahlbruch}}, \bibinfo {author} {\bibfnamefont {M.}~\bibnamefont {Mehmet}},
  \bibinfo {author} {\bibfnamefont {S.}~\bibnamefont {Chelkowski}}, \bibinfo
  {author} {\bibfnamefont {B.}~\bibnamefont {Hage}}, \bibinfo {author}
  {\bibfnamefont {A.}~\bibnamefont {Franzen}}, \bibinfo {author} {\bibfnamefont
  {N.}~\bibnamefont {Lastzka}}, \bibinfo {author} {\bibfnamefont
  {S.}~\bibnamefont {Go{\ss}ler}}, \bibinfo {author} {\bibfnamefont
  {K.}~\bibnamefont {Danzmann}}, \ and\ \bibinfo {author} {\bibfnamefont
  {R.}~\bibnamefont {Schnabel}},\ }\href@noop {} {\bibfield  {journal}
  {\bibinfo  {journal} {Physical Review Letters}\ }\textbf {\bibinfo {volume}
  {100}},\ \bibinfo {pages} {033602} (\bibinfo {year} {2008})}\BibitemShut
  {NoStop}%
\bibitem [{\citenamefont {Schnabel}(2017)}]{2017PhR...684....1S}%
  \BibitemOpen
  \bibfield  {author} {\bibinfo {author} {\bibfnamefont {R.}~\bibnamefont
  {Schnabel}},\ }\href@noop {} {\bibfield  {journal} {\bibinfo  {journal}
  {Physics Reports}\ }\textbf {\bibinfo {volume} {684}},\ \bibinfo {pages} {1}
  (\bibinfo {year} {2017})}\BibitemShut {NoStop}%
\bibitem [{\citenamefont {F{\"o}rtsch}\ \emph {et~al.}(2013)\citenamefont
  {F{\"o}rtsch}, \citenamefont {F{\"u}rst}, \citenamefont {Wittmann},
  \citenamefont {Strekalov}, \citenamefont {Aiello}, \citenamefont {Chekhova},
  \citenamefont {Silberhorn}, \citenamefont {Leuchs},\ and\ \citenamefont
  {Marquardt}}]{2013NatCo...4.1818F}%
  \BibitemOpen
  \bibfield  {author} {\bibinfo {author} {\bibfnamefont {M.}~\bibnamefont
  {F{\"o}rtsch}}, \bibinfo {author} {\bibfnamefont {J.~U.}\ \bibnamefont
  {F{\"u}rst}}, \bibinfo {author} {\bibfnamefont {C.}~\bibnamefont {Wittmann}},
  \bibinfo {author} {\bibfnamefont {D.}~\bibnamefont {Strekalov}}, \bibinfo
  {author} {\bibfnamefont {A.}~\bibnamefont {Aiello}}, \bibinfo {author}
  {\bibfnamefont {M.~V.}\ \bibnamefont {Chekhova}}, \bibinfo {author}
  {\bibfnamefont {C.}~\bibnamefont {Silberhorn}}, \bibinfo {author}
  {\bibfnamefont {G.}~\bibnamefont {Leuchs}}, \ and\ \bibinfo {author}
  {\bibfnamefont {C.}~\bibnamefont {Marquardt}},\ }\href@noop {} {\bibfield
  {journal} {\bibinfo  {journal} {Nature Communications}\ }\textbf {\bibinfo
  {volume} {4}},\ \bibinfo {pages} {1818} (\bibinfo {year} {2013})}\BibitemShut
  {NoStop}%
\bibitem [{\citenamefont {Pirandola}\ \emph {et~al.}(2015)\citenamefont
  {Pirandola}, \citenamefont {Eisert}, \citenamefont {Weedbrook}, \citenamefont
  {Furusawa},\ and\ \citenamefont {Braunstein}}]{2015NaPho...9..641P}%
  \BibitemOpen
  \bibfield  {author} {\bibinfo {author} {\bibfnamefont {S.}~\bibnamefont
  {Pirandola}}, \bibinfo {author} {\bibfnamefont {J.}~\bibnamefont {Eisert}},
  \bibinfo {author} {\bibfnamefont {C.}~\bibnamefont {Weedbrook}}, \bibinfo
  {author} {\bibfnamefont {A.}~\bibnamefont {Furusawa}}, \ and\ \bibinfo
  {author} {\bibfnamefont {S.~L.}\ \bibnamefont {Braunstein}},\ }\href@noop {}
  {\bibfield  {journal} {\bibinfo  {journal} {Nature Photonics}\ }\textbf
  {\bibinfo {volume} {9}},\ \bibinfo {pages} {641} (\bibinfo {year}
  {2015})}\BibitemShut {NoStop}%
\bibitem [{\citenamefont {Menzel}\ \emph {et~al.}(2012)\citenamefont {Menzel},
  \citenamefont {Di~Candia}, \citenamefont {Deppe}, \citenamefont {Eder},
  \citenamefont {Zhong}, \citenamefont {Ihmig}, \citenamefont {Haeberlein},
  \citenamefont {Baust}, \citenamefont {Hoffmann}, \citenamefont {Ballester},
  \citenamefont {Inomata}, \citenamefont {Yamamoto}, \citenamefont {Nakamura},
  \citenamefont {Solano}, \citenamefont {Marx},\ and\ \citenamefont
  {Gross}}]{2012PhRvL.109y0502M}%
  \BibitemOpen
  \bibfield  {author} {\bibinfo {author} {\bibfnamefont {E.~P.}\ \bibnamefont
  {Menzel}}, \bibinfo {author} {\bibfnamefont {R.}~\bibnamefont {Di~Candia}},
  \bibinfo {author} {\bibfnamefont {F.}~\bibnamefont {Deppe}}, \bibinfo
  {author} {\bibfnamefont {P.}~\bibnamefont {Eder}}, \bibinfo {author}
  {\bibfnamefont {L.}~\bibnamefont {Zhong}}, \bibinfo {author} {\bibfnamefont
  {M.}~\bibnamefont {Ihmig}}, \bibinfo {author} {\bibfnamefont
  {M.}~\bibnamefont {Haeberlein}}, \bibinfo {author} {\bibfnamefont
  {A.}~\bibnamefont {Baust}}, \bibinfo {author} {\bibfnamefont
  {E.}~\bibnamefont {Hoffmann}}, \bibinfo {author} {\bibfnamefont
  {D.}~\bibnamefont {Ballester}}, \bibinfo {author} {\bibfnamefont
  {K.}~\bibnamefont {Inomata}}, \bibinfo {author} {\bibfnamefont
  {T.}~\bibnamefont {Yamamoto}}, \bibinfo {author} {\bibfnamefont
  {Y.}~\bibnamefont {Nakamura}}, \bibinfo {author} {\bibfnamefont
  {E.}~\bibnamefont {Solano}}, \bibinfo {author} {\bibfnamefont
  {A.}~\bibnamefont {Marx}}, \ and\ \bibinfo {author} {\bibfnamefont
  {R.}~\bibnamefont {Gross}},\ }\href@noop {} {\bibfield  {journal} {\bibinfo
  {journal} {Physical Review Letters}\ }\textbf {\bibinfo {volume} {109}},\
  \bibinfo {pages} {250502} (\bibinfo {year} {2012})}\BibitemShut {NoStop}%
\bibitem [{\citenamefont {Eichler}\ \emph {et~al.}(2014)\citenamefont
  {Eichler}, \citenamefont {Salath{\'e}}, \citenamefont {Mlynek}, \citenamefont
  {Schmidt},\ and\ \citenamefont {Wallraff}}]{2014PhRvL.113k0502E}%
  \BibitemOpen
  \bibfield  {author} {\bibinfo {author} {\bibfnamefont {C.}~\bibnamefont
  {Eichler}}, \bibinfo {author} {\bibfnamefont {Y.}~\bibnamefont
  {Salath{\'e}}}, \bibinfo {author} {\bibfnamefont {J.}~\bibnamefont {Mlynek}},
  \bibinfo {author} {\bibfnamefont {S.}~\bibnamefont {Schmidt}}, \ and\
  \bibinfo {author} {\bibfnamefont {A.}~\bibnamefont {Wallraff}},\ }\href@noop
  {} {\bibfield  {journal} {\bibinfo  {journal} {Physical Review Letters}\
  }\textbf {\bibinfo {volume} {113}},\ \bibinfo {pages} {110502} (\bibinfo
  {year} {2014})}\BibitemShut {NoStop}%
\bibitem [{\citenamefont {Fedorov}\ \emph {et~al.}(2016)\citenamefont
  {Fedorov}, \citenamefont {Zhong}, \citenamefont {Pogorzalek}, \citenamefont
  {Eder}, \citenamefont {Fischer}, \citenamefont {Goetz}, \citenamefont {Xie},
  \citenamefont {Wulschner}, \citenamefont {Inomata}, \citenamefont {Yamamoto},
  \citenamefont {Nakamura}, \citenamefont {Di~Candia}, \citenamefont
  {Las~Heras}, \citenamefont {Sanz}, \citenamefont {Solano}, \citenamefont
  {Menzel}, \citenamefont {Deppe}, \citenamefont {Marx},\ and\ \citenamefont
  {Gross}}]{2016PhRvL.117b0502F}%
  \BibitemOpen
  \bibfield  {author} {\bibinfo {author} {\bibfnamefont {K.~G.}\ \bibnamefont
  {Fedorov}}, \bibinfo {author} {\bibfnamefont {L.}~\bibnamefont {Zhong}},
  \bibinfo {author} {\bibfnamefont {S.}~\bibnamefont {Pogorzalek}}, \bibinfo
  {author} {\bibfnamefont {P.}~\bibnamefont {Eder}}, \bibinfo {author}
  {\bibfnamefont {M.}~\bibnamefont {Fischer}}, \bibinfo {author} {\bibfnamefont
  {J.}~\bibnamefont {Goetz}}, \bibinfo {author} {\bibfnamefont
  {E.}~\bibnamefont {Xie}}, \bibinfo {author} {\bibfnamefont {F.}~\bibnamefont
  {Wulschner}}, \bibinfo {author} {\bibfnamefont {K.}~\bibnamefont {Inomata}},
  \bibinfo {author} {\bibfnamefont {T.}~\bibnamefont {Yamamoto}}, \bibinfo
  {author} {\bibfnamefont {Y.}~\bibnamefont {Nakamura}}, \bibinfo {author}
  {\bibfnamefont {R.}~\bibnamefont {Di~Candia}}, \bibinfo {author}
  {\bibfnamefont {U.}~\bibnamefont {Las~Heras}}, \bibinfo {author}
  {\bibfnamefont {M.}~\bibnamefont {Sanz}}, \bibinfo {author} {\bibfnamefont
  {E.}~\bibnamefont {Solano}}, \bibinfo {author} {\bibfnamefont {E.~P.}\
  \bibnamefont {Menzel}}, \bibinfo {author} {\bibfnamefont {F.}~\bibnamefont
  {Deppe}}, \bibinfo {author} {\bibfnamefont {A.}~\bibnamefont {Marx}}, \ and\
  \bibinfo {author} {\bibfnamefont {R.}~\bibnamefont {Gross}},\ }\href@noop {}
  {\bibfield  {journal} {\bibinfo  {journal} {Physical Review Letters}\
  }\textbf {\bibinfo {volume} {117}},\ \bibinfo {pages} {020502} (\bibinfo
  {year} {2016})}\BibitemShut {NoStop}%
\bibitem [{\citenamefont {Bienfait}\ \emph {et~al.}(2017)\citenamefont
  {Bienfait}, \citenamefont {Campagne-Ibarcq}, \citenamefont {Kiilerich},
  \citenamefont {Zhou}, \citenamefont {Probst}, \citenamefont {Pla},
  \citenamefont {Schenkel}, \citenamefont {Vion}, \citenamefont {Esteve},
  \citenamefont {Morton}, \citenamefont {Moelmer},\ and\ \citenamefont
  {Bertet}}]{2017PhRvX...7d1011B}%
  \BibitemOpen
  \bibfield  {author} {\bibinfo {author} {\bibfnamefont {A.}~\bibnamefont
  {Bienfait}}, \bibinfo {author} {\bibfnamefont {P.}~\bibnamefont
  {Campagne-Ibarcq}}, \bibinfo {author} {\bibfnamefont {A.~H.}\ \bibnamefont
  {Kiilerich}}, \bibinfo {author} {\bibfnamefont {X.}~\bibnamefont {Zhou}},
  \bibinfo {author} {\bibfnamefont {S.}~\bibnamefont {Probst}}, \bibinfo
  {author} {\bibfnamefont {J.~J.}\ \bibnamefont {Pla}}, \bibinfo {author}
  {\bibfnamefont {T.}~\bibnamefont {Schenkel}}, \bibinfo {author}
  {\bibfnamefont {D.}~\bibnamefont {Vion}}, \bibinfo {author} {\bibfnamefont
  {D.}~\bibnamefont {Esteve}}, \bibinfo {author} {\bibfnamefont {J.~J.~L.}\
  \bibnamefont {Morton}}, \bibinfo {author} {\bibfnamefont {K.}~\bibnamefont
  {Moelmer}}, \ and\ \bibinfo {author} {\bibfnamefont {P.}~\bibnamefont
  {Bertet}},\ }\href@noop {} {\bibfield  {journal} {\bibinfo  {journal}
  {Physical Review X}\ }\textbf {\bibinfo {volume} {7}},\ \bibinfo {pages}
  {041011} (\bibinfo {year} {2017})}\BibitemShut {NoStop}%
\bibitem [{\citenamefont {Fuwa}\ \emph {et~al.}(2015)\citenamefont {Fuwa},
  \citenamefont {Takeda}, \citenamefont {Zwierz}, \citenamefont {Wiseman},\
  and\ \citenamefont {Furusawa}}]{2015NatCo...6E6665F}%
  \BibitemOpen
  \bibfield  {author} {\bibinfo {author} {\bibfnamefont {M.}~\bibnamefont
  {Fuwa}}, \bibinfo {author} {\bibfnamefont {S.}~\bibnamefont {Takeda}},
  \bibinfo {author} {\bibfnamefont {M.}~\bibnamefont {Zwierz}}, \bibinfo
  {author} {\bibfnamefont {H.~M.}\ \bibnamefont {Wiseman}}, \ and\ \bibinfo
  {author} {\bibfnamefont {A.}~\bibnamefont {Furusawa}},\ }\href@noop {}
  {\bibfield  {journal} {\bibinfo  {journal} {Nature Communications}\ }\textbf
  {\bibinfo {volume} {6}},\ \bibinfo {pages} {6665} (\bibinfo {year}
  {2015})}\BibitemShut {NoStop}%
\bibitem [{\citenamefont {Tsang}(2010)}]{2010PhRvA..81f3837T}%
  \BibitemOpen
  \bibfield  {author} {\bibinfo {author} {\bibfnamefont {M.}~\bibnamefont
  {Tsang}},\ }\href@noop {} {\bibfield  {journal} {\bibinfo  {journal}
  {Physical Review A}\ }\textbf {\bibinfo {volume} {81}},\ \bibinfo {pages}
  {063837} (\bibinfo {year} {2010})}\BibitemShut {NoStop}%
\bibitem [{\citenamefont {Tsang}(2011)}]{2011PhRvA..84d3845T}%
  \BibitemOpen
  \bibfield  {author} {\bibinfo {author} {\bibfnamefont {M.}~\bibnamefont
  {Tsang}},\ }\href@noop {} {\bibfield  {journal} {\bibinfo  {journal}
  {Physical Review A}\ }\textbf {\bibinfo {volume} {84}},\ \bibinfo {pages}
  {043845} (\bibinfo {year} {2011})}\BibitemShut {NoStop}%
\bibitem [{\citenamefont {Rueda}\ \emph {et~al.}(2016)\citenamefont {Rueda},
  \citenamefont {Sedlmeir}, \citenamefont {Collodo}, \citenamefont {Vogl},
  \citenamefont {Stiller}, \citenamefont {Schunk}, \citenamefont {Strekalov},
  \citenamefont {Marquardt}, \citenamefont {Fink}, \citenamefont {Painter},
  \citenamefont {Leuchs},\ and\ \citenamefont {Schwefel}}]{Rueda:2016jg}%
  \BibitemOpen
  \bibfield  {author} {\bibinfo {author} {\bibfnamefont {A.}~\bibnamefont
  {Rueda}}, \bibinfo {author} {\bibfnamefont {F.}~\bibnamefont {Sedlmeir}},
  \bibinfo {author} {\bibfnamefont {M.~C.}\ \bibnamefont {Collodo}}, \bibinfo
  {author} {\bibfnamefont {U.}~\bibnamefont {Vogl}}, \bibinfo {author}
  {\bibfnamefont {B.}~\bibnamefont {Stiller}}, \bibinfo {author} {\bibfnamefont
  {G.}~\bibnamefont {Schunk}}, \bibinfo {author} {\bibfnamefont {D.~V.}\
  \bibnamefont {Strekalov}}, \bibinfo {author} {\bibfnamefont {C.}~\bibnamefont
  {Marquardt}}, \bibinfo {author} {\bibfnamefont {J.~M.}\ \bibnamefont {Fink}},
  \bibinfo {author} {\bibfnamefont {O.}~\bibnamefont {Painter}}, \bibinfo
  {author} {\bibfnamefont {G.}~\bibnamefont {Leuchs}}, \ and\ \bibinfo {author}
  {\bibfnamefont {H.~G.~L.}\ \bibnamefont {Schwefel}},\ }\href@noop {}
  {\bibfield  {journal} {\bibinfo  {journal} {Optica}\ }\textbf {\bibinfo
  {volume} {3}},\ \bibinfo {pages} {597} (\bibinfo {year} {2016})}\BibitemShut
  {NoStop}%
\bibitem [{\citenamefont {Zhang}\ \emph {et~al.}(2016)\citenamefont {Zhang},
  \citenamefont {Zou}, \citenamefont {Jiang},\ and\ \citenamefont
  {Tang}}]{2016SciA....2E1286Z}%
  \BibitemOpen
  \bibfield  {author} {\bibinfo {author} {\bibfnamefont {X.}~\bibnamefont
  {Zhang}}, \bibinfo {author} {\bibfnamefont {C.-L.}\ \bibnamefont {Zou}},
  \bibinfo {author} {\bibfnamefont {L.}~\bibnamefont {Jiang}}, \ and\ \bibinfo
  {author} {\bibfnamefont {H.~X.}\ \bibnamefont {Tang}},\ }\href@noop {}
  {\bibfield  {journal} {\bibinfo  {journal} {Science Advances}\ }\textbf
  {\bibinfo {volume} {2}},\ \bibinfo {pages} {e1501286} (\bibinfo {year}
  {2016})}\BibitemShut {NoStop}%
\bibitem [{\citenamefont {Asjad}\ \emph {et~al.}(2019)\citenamefont {Asjad},
  \citenamefont {Abari}, \citenamefont {Zippilli},\ and\ \citenamefont
  {Vitali}}]{Asjad:2019}%
  \BibitemOpen
  \bibfield  {author} {\bibinfo {author} {\bibfnamefont {M.}~\bibnamefont
  {Asjad}}, \bibinfo {author} {\bibfnamefont {N.~E.}\ \bibnamefont {Abari}},
  \bibinfo {author} {\bibfnamefont {S.}~\bibnamefont {Zippilli}}, \ and\
  \bibinfo {author} {\bibfnamefont {D.}~\bibnamefont {Vitali}},\ }\href@noop {}
  {\bibfield  {journal} {\bibinfo  {journal} {Optics Express}\ }\textbf
  {\bibinfo {volume} {27}},\ \bibinfo {pages} {32427} (\bibinfo {year}
  {2019})}\BibitemShut {NoStop}%
\bibitem [{\citenamefont {Gan}\ \emph {et~al.}(2019)\citenamefont {Gan},
  \citenamefont {Liu}, \citenamefont {Lu}, \citenamefont {Wang}, \citenamefont
  {Tey},\ and\ \citenamefont {You}}]{Gan:2019ef}%
  \BibitemOpen
  \bibfield  {author} {\bibinfo {author} {\bibfnamefont {J.~H.}\ \bibnamefont
  {Gan}}, \bibinfo {author} {\bibfnamefont {Y.~C.}\ \bibnamefont {Liu}},
  \bibinfo {author} {\bibfnamefont {C.}~\bibnamefont {Lu}}, \bibinfo {author}
  {\bibfnamefont {X.}~\bibnamefont {Wang}}, \bibinfo {author} {\bibfnamefont
  {M.~K.}\ \bibnamefont {Tey}}, \ and\ \bibinfo {author} {\bibfnamefont
  {L.}~\bibnamefont {You}},\ }\href@noop {} {\bibfield  {journal} {\bibinfo
  {journal} {Laser {\&} Photonics Reviews}\ }\textbf {\bibinfo {volume} {13}},\
  \bibinfo {pages} {1900120} (\bibinfo {year} {2019})}\BibitemShut {NoStop}%
\bibitem [{\citenamefont {Wang}\ \emph {et~al.}(2015)\citenamefont {Wang},
  \citenamefont {Chesi},\ and\ \citenamefont {Clerk}}]{2015PhRvA..91a3807W}%
  \BibitemOpen
  \bibfield  {author} {\bibinfo {author} {\bibfnamefont {Y.-D.}\ \bibnamefont
  {Wang}}, \bibinfo {author} {\bibfnamefont {S.}~\bibnamefont {Chesi}}, \ and\
  \bibinfo {author} {\bibfnamefont {A.~A.}\ \bibnamefont {Clerk}},\ }\href@noop
  {} {\bibfield  {journal} {\bibinfo  {journal} {Physical Review A}\ }\textbf
  {\bibinfo {volume} {91}},\ \bibinfo {pages} {013807} (\bibinfo {year}
  {2015})}\BibitemShut {NoStop}%
\bibitem [{\citenamefont {DeJesus}\ and\ \citenamefont
  {Kaufman}(1987)}]{1987PhRvA..35.5288D}%
  \BibitemOpen
  \bibfield  {author} {\bibinfo {author} {\bibfnamefont {E.~X.}\ \bibnamefont
  {DeJesus}}\ and\ \bibinfo {author} {\bibfnamefont {C.}~\bibnamefont
  {Kaufman}},\ }\href@noop {} {\bibfield  {journal} {\bibinfo  {journal}
  {Physical Review A}\ }\textbf {\bibinfo {volume} {35}},\ \bibinfo {pages}
  {5288} (\bibinfo {year} {1987})}\BibitemShut {NoStop}%
\bibitem [{\citenamefont {Peterson}\ \emph {et~al.}(2019)\citenamefont
  {Peterson}, \citenamefont {Kotler}, \citenamefont {Lecocq}, \citenamefont
  {Cicak}, \citenamefont {Jin}, \citenamefont {Simmonds}, \citenamefont
  {Aumentado},\ and\ \citenamefont {Teufel}}]{Peterson:2019}%
  \BibitemOpen
  \bibfield  {author} {\bibinfo {author} {\bibfnamefont {G.~A.}\ \bibnamefont
  {Peterson}}, \bibinfo {author} {\bibfnamefont {S.}~\bibnamefont {Kotler}},
  \bibinfo {author} {\bibfnamefont {F.}~\bibnamefont {Lecocq}}, \bibinfo
  {author} {\bibfnamefont {K.}~\bibnamefont {Cicak}}, \bibinfo {author}
  {\bibfnamefont {X.~Y.}\ \bibnamefont {Jin}}, \bibinfo {author} {\bibfnamefont
  {R.~W.}\ \bibnamefont {Simmonds}}, \bibinfo {author} {\bibfnamefont
  {J.}~\bibnamefont {Aumentado}}, \ and\ \bibinfo {author} {\bibfnamefont
  {J.~D.}\ \bibnamefont {Teufel}},\ }\href@noop {} {\bibfield  {journal}
  {\bibinfo  {journal} {Physical Review Letters}\ }\textbf {\bibinfo {volume}
  {123}},\ \bibinfo {pages} {247701} (\bibinfo {year} {2019})}\BibitemShut
  {NoStop}%
\bibitem [{\citenamefont {Filip}(2009)}]{2009PhRvA..80b2304F}%
  \BibitemOpen
  \bibfield  {author} {\bibinfo {author} {\bibfnamefont {R.}~\bibnamefont
  {Filip}},\ }\href@noop {} {\bibfield  {journal} {\bibinfo  {journal}
  {Physical Review A}\ }\textbf {\bibinfo {volume} {80}},\ \bibinfo {pages}
  {022304} (\bibinfo {year} {2009})}\BibitemShut {NoStop}%
\bibitem [{\citenamefont {Zhang}\ \emph {et~al.}(2018)\citenamefont {Zhang},
  \citenamefont {Zou},\ and\ \citenamefont {Jiang}}]{2018PhRvL.120b0502Z}%
  \BibitemOpen
  \bibfield  {author} {\bibinfo {author} {\bibfnamefont {M.}~\bibnamefont
  {Zhang}}, \bibinfo {author} {\bibfnamefont {C.-L.}\ \bibnamefont {Zou}}, \
  and\ \bibinfo {author} {\bibfnamefont {L.}~\bibnamefont {Jiang}},\
  }\href@noop {} {\bibfield  {journal} {\bibinfo  {journal} {Physical Review
  Letters}\ }\textbf {\bibinfo {volume} {120}},\ \bibinfo {pages} {020502}
  (\bibinfo {year} {2018})}\BibitemShut {NoStop}%
\end{thebibliography}%

\end{document}